\documentclass[useAMS,usenatbib]{mn2e}

\newcommand{\FIG}[1]{}
\newcommand{\superscript}[1]{\ensuremath{^\textrm{#1}}}

\usepackage{amssymb}
\usepackage{amsmath}
\usepackage{subfigure}
\usepackage{textcomp}
\usepackage{graphicx}

\title{On the thermal sensitivity of binary formation in collapsing molecular clouds}

\author[R. Riaz, S. Z. Farooqui and S. Vanaverbeke]{R. Riaz$^{1}$\thanks{E-mail:
rafilriaz@yahoo.com} 
S. Z. Farooqui$^{2}$\thanks{E-mail: drsuzaki@hotmail.com}
S. Vanaverbeke$^{3,4}$\thanks{E-mail:
Sigfried.Vanaverbeke@kuleuven-kortrijk.be} \\
$^{1}$Institute of Space and Planetary Astrophysics (ISPA),
University of Karachi, Karachi 75270, Pakistan \\
$^{2}$Faculty of Engineering Sciences, National University of Science
Technology, Islamabad, Pakistan \\
$^{3}$Centre for mathematical Plasma-Astrophysics, Department of Mathematics, KU Leuven, Celestijnenlaan
200B, 3001 Heverlee, Belgium \\
$^{4}$Katholieke Universiteit Leuven Campus Kortrijk, E.-Sabbelaan 53, 8500 Kortrijk,Belgium}

\begin{document}

\date{Accepted - Received -}

\pagerange{\pageref{firstpage}--\pageref{lastpage}} \pubyear{2014}

\maketitle

\label{firstpage}

\begin{abstract}

We report the results of a numerical study on the initial formation stages of low-mass protostellar binary systems. We determine the separation of protostellar binaries formed as a function of the initial thermal state by varying the initial temperature in a slightly modified version of the Burkert and Bodenheimer collapse test. We find that the outcome is highly sensitive to both the initial temperature of the cloud and the initial amplitude of azimuthal density perturbation A. For A=10 $\%$, variations of only 1 unit Kelvin below 10 K lead to changes of up to 100 AU ( i.e.  of order 30 \%) in the instantaneous separation, whereas for this small A the initial temperatures above 10 K yield, instead of a binary, a single low-mass fragment that never reaches protostellar densities. Protostellar binaries, however, do emerge when the perturbation amplitude is increased from 10 $\%$ to 25 $\%$.
We also investigate the impact of the critical density which governs the transition from isothermal to adiabatic thermodynamic behaviour of the collapsing gas. We find that the critical density not only affects the overall structural evolution of the gas envelope, but also the size of the rotating disk structures formed during collapse as well as the number of protostellar fragments resulting from the final fragmentation of the disks. This mechanism can give rise to young protostellar objects constituting bound multiple stellar systems.

\end{abstract}


\section{Introduction}

Molecular clouds are believed to be among the coolest regions in a galaxy where low-mass star formation can take place. Because star formation involves an enormous range of complex physical processes, it is difficult to construct exhaustive theoretical models which include every aspect of the problem (\cite{Krumholtz}).  
It is generally found that systems consisting of multiple protostars emerge as a result of the gravitational collapse of rotating molecular cloud cores (\cite{Raghavan,Janson}).
A significant fraction of these multiple star systems are binary systems. 
In fact, infrared surveys of nearby star formation regions reveal that around $~75\%$ of the newly formed protostars are part of multiple stellar systems (\cite{Duchêne}). Furthermore, an interesting population of loosely bound very low-mass (VLM) binaries with large separation $d$ ($d >$ 100 AU) is also found in very young star forming regions( e.g. \cite{Béjar,Bouy,Luhman}). 

Theories which try to explain the formation processes of binary star systems include disk fragmentation, fission of a protostar(s), and dynamical capture of protostars into bound systems (\cite{Tobin,Moeckel}). 
Among these possible scenarios, the disk fragmentation model is the subject of the investigation reported in this paper. The basic purpose of this paper is to reexamine numerically certain aspects of the gravitational collapse and the subsequent fragmentation of rotating  molecular cloud cores. We focus on the dependence of disk fragmentation as a low-mass binary and multiple star formation mechanism on the initial temperature as well as the critical density which marks the transition from isothermal to adiabatic thermodynamic behaviour of the collapsing gas. 

A significant number of simulation results have so far been reported in the literature in which binary systems emerging from disk fragmentation have been investigated (\cite{Fumitaka,Machida,Forgan}). The availability of advanced computational resources and improved versions of state-of-the-art star formation codes have significantly contributed to our understanding of the physical processes involved in star formation (\cite{Frank,Loughnane,Hubber,Duffin}). Among the recent efforts, we mention numerical models of the collapse of molecular clouds to protostellar densities from an initial Gaussian and Plummer density distribution (\cite{Gomez-Ramirez}). The impact of magnetic fields on 
protostellar collapse and fragmentation has also been investigated in detail (for example, \cite{Bürzle,Donati,Peters,Boss}). 
 
Based on the results of his collapse calculations including radiative transfer in the SPH framework, \cite{Stamatellos} found that for densities below 10\superscript{-13} g/cm\superscript{3}, the temperature of the gas can be approximated by the following expression:

\begin{equation} \label{Tdependence}
T=5 K\left( \frac{\rho}{10\superscript{-13} g/cm\superscript{3}}\right)^{0.08}.
\end{equation}

The above relation describes the slow heating process in collapsing cores within the density range  10\superscript{-18} g/cm\superscript{3} $< \rho <$ 10\superscript{-13} g/cm\superscript{3}. Within this range the effective equation of state of the gas is almost isothermal. Beyond this density range the temperature creeps up much more rapidly due to a rise in the opacity of the gas. As a result, the model equation \ref{Tdependence} can not be used any more to estimate the temperature of the cloud.

Furthermore, in a recent paper, \cite{Launhardt} revealed the decisive role which is played by physical quantities such as the initial temperature and the initial density structure of the cloud in determining the final end products of gravitationally collapsing cores. Molecular gas with a slightly enhanced metallicity compared to the metallicity which is normally found in the solar neighborhood could provide an efficient cooling mechanism in collapsing clouds. Metal rich gas allows molecular cloud cores to remain isothermal for a longer period of time because a cloud with a relatively higher metallicity would be subject to more efficient radiative cooling, hence prolonging its isothermal state during collapse compared to clouds in which metal poor conditions prevail (see for example \cite{Omukai}). To mimic the effect of varying the metallicity, we have used a barotropic equation of state and analyzed the evolution of rotating collapsing molecular cloud cores by varying the value of the transition density at which the gas switches its thermodynamic behaviour from isothermal to adiabatic. A similar strategy has been adopted by \cite{Gomez-Ramirez}. The main difference of their setup with our work is that we consider initial conditions with uniform density rather than Gaussian and Plummer models.
On the other hand, observations of molecular cloud cores have revealed values for the lowest temperatures down to at least 8K with typical values of 10K (\cite{BensonMyers,WardThompson}).Therefore we explored a slightly unorthodox but yet possible range of initial temperatures in star-forming molecular clouds between 8K to 12K, keeping in mind that several physical mechanisms can cause prestellar gas to heat up or to cool down from the commonly adopted value of 10 K.  For example, cosmic rays can ionize the medium causing the temperature to climb up from 10K to a certain limit or line radiation from molecules can cool down the gas below this value(\cite{Padovani,Indriolo,Wiener}). This leads us to consider the possible role of the initial thermal state of star-forming clouds in determining the properties of binary protostar systems. We focus on the relation between the binary separation of the protostars formed through gravitational collapse and the initial temperature of the cloud, as well as the possible emergence of protostellar objects as a result of secondary fragmentation.  

The numerical models of cloud collapse recently reported in (\cite{Gomez-Ramirez}) demonstrate  that there exists a tendency for models with initial Plummer and Gaussian density distributions to switch from the isothermal to the adiabatic regime at slightly lower critical densities. Models with a lower critical density show enhanced fragmentation compared to models with higher critical densities. 
Keeping this in mind, we also examine the relation between the critical density and  
the number of fragments resulting from the gravitational collapse. We also present an analysis of the extreme values of density and temperature within the fragments.
The structure of the paper is as follows. In section 2, we provide details on the SPH code GRADSPH which has been used in this work. We describe the setup of the numerical models in section 3.  Section 4 gives an overview of the results of our simulations. Finally, our conclusions are presented in section 5. 

\section{Numerical Method}

For the hydrodynamical models of molecular could collapse considered in this work we use GRADSPH, a tree-based, parallel particle-based hydrodynamics code based on the Smoothed Particle Hydrodynamics (SPH) method written in FORTRAN 90(\cite{GRADSPH}). The code has several features which enable us to model gravitational cloud collapse, including the treatment of gravitational forces using a tree-based gravity (TCG) method combined with a variable gravitational softening length formalism(for details see \cite{GRADSPH} and \cite{PriceMon}). The SPH formalism implemented in GRADSPH is based on deriving the SPH equations from a variational principle(\cite{Pricereview}). The reliability of the code has been established through an extensive series of tests on standard benchmark problems and(\cite{GRADSPH}) and the code has recently been extended to MHD(\cite{GRADSPMHD}). 

According to (\cite{Pricereview}),(see also \cite{Gingold}),the basic equations of motion used for evolving the SPH particles in GRADSPH are given as follows:

\begin{equation} \label{GRADSPHeom}
\begin{split}
& \frac{d\vec{v}_{i} }{dt} = \\ & 
-\sum _{j=1}^{N} m_{j} \left[\frac{P_{i} }{\rho _{i}^{2} \Omega_{i}}\nabla _{i} W(r_{ij} ,h_{i})
+ \frac{P_{j} }{\rho_{j}^{2} \Omega_{j} } \nabla _{i} W(r_{ij} ,h_{j}) \right],
\end{split}
\end{equation}

in which the coefficients $\Omega_{i}$ are defined as
\begin{equation} \label{omega}
\Omega_{i}=1-\frac{\partial h_{i}}{\partial \rho_{i}}\sum_{j=1}^{N} m_{j} \frac{\partial W(r_{ij},h_{j})}{\partial h_{j}} 
\end{equation}

In the above equations, $N$ is the total number of particles in the simulation, $P_{i}$ and $\rho_{i}$ denote the density and pressure of a particle labeled $i$, $h_{i}$ is its smoothing length and $m_{i}$ denotes its mass.
The position and velocity vectors for the particles are denoted by $\vec{r}_{i}$ and $\vec{v}_{i}$, respectively, and $\vec{r}_{ij}=\vec{r}_{i}-\vec{r}_{j}$. 
The coefficients $\Omega_{i}$ self-consistently take into account the effect of a variable smoothing length (\cite{Pricereview}). In this paper, we close the system of equations by specifying a barotropic equation of state $P=P\left(\rho\right)$ for the gas which will be specified in section 3. In this way we eliminate the need to solve an energy equation. The code uses time-dependent artificial viscosity terms in order to capture shock waves and prevent particle penetration. In order to maintain hydrodynamic stability, we use the signal-velocity approach introduced by Price and Monaghan (\cite{PriceMon}) to calculate the artificial viscosity terms and  the Courant time step. The Courant number is set to 0.1 in our simulations. This choice is found to be sufficient to maintain stability in our calculations. In the artificial viscosity terms, we set the parameters $C_{a}$ and $\alpha_{min}$ in Eqns. (34) and  (35) in (\cite{GRADSPH}) to 0.2 and 0.5, respectively.

The particle densities are computed from the standard SPH summation equation (\cite{Pricereview,GRADSPH})
by summing the contribution from all the particles which overlap with the position of particle $i$, 
using a weighting function $W\left(\vec{r}_{i}-\vec{r}_{j},h_{i}\right)$:
\begin{equation} \label{density}
\rho _{i} =\sum _{j=1}^{N} m_{j} W\left(\vec{r}_{i}-\vec{r}_{j},h_{i}\right).
\end{equation}
In the above expression, $W\left(\vec{r}_{i}-\vec{r}_{j},h_{i}\right)$ is a smooth differentiable function, referred to as the smoothing kernel or the interpolating kernel. GRADSPH uses the standard M4-kernel or cubic spline kernel (\cite{Pricereview}).
The smoothing length is updated at each time step by iteratively solving the following equation for each particle:(\cite{Pricereview}): 

\begin{equation} \label{smoothinglength}
\frac{4\pi}{3}\left(2 h_{i}\right)^{3}\rho_{i}=m_{i}N_{opt}, 
\end{equation}

in which $N_{opt}$ is the number of neighbours contained within the smoothing sphere of each particle. 
$N_{opt}$ is set to 50 in our calculations.
In this way we ensure that a constant mass is contained within the smoothing sphere for every SPH particle 
so that the code adapts its resolution to keep track of density changes that may occur during the dynamical evolution of the fluid. The system of ordinary differential equations \ref{GRADSPHeom}, which updates the positions and velocities of the particles, is solved using a predictor-corrector scheme combined with an individual particle time stepping method.

As mentioned before, the self-gravity of the gas is treated using the TCG method. We compute the gravitational acceleration of each particle using a Barnes-Hut tree algorithm. The opening angle for the tree is an important parameter which we set to $\theta$ = 0.7. 
The same Barnes-Hut tree algorithm is also used to update the list of neighbours of the SPH particles. We use the cubic spline kernel to soften gravitational forces and include the correction terms derived by (\cite{PriceMon}) to ensure the conservation of energy when dealing with variable particle smoothing lengths. The gravitational acceleration of particle $i$ is thus given by 

\begin{equation} \label{gravaccel}
\begin{split}
& \vec{g}_{i}=-G \sum_{j=1}^{N}\left[\Phi^{'}\left(r_{ij},h_{i}\right)+   
\Phi^{'}\left(r_{ij},h_{j}\right)\right]\vec{e}_{ij}- \\ &
\frac{G}{2}\sum_{j=1}^{N}\left[\frac{\zeta_{i}}{\Omega_{i}}\nabla_{i}W\left(r_{ij},h_{i}\right)
+ \frac{\zeta_{j}}{\Omega_{j}}\nabla_{i}W\left(r_{ij},h_{i}\right)\right]
\end{split}
\end{equation} 

in which the quantities $\zeta_{i}$ are defined as

\begin{equation} \label{zeta}
\zeta_{i}=\frac{\partial h_{i}}{\partial \rho_{i}}\sum_{j=1}^{N} m_{j} \frac{\partial \Phi(r_{ij},h_{i})}{\partial h_{i}},
\end{equation}

$\Phi\left(r,h\right)$ is the softened gravitational potential of a particle, and $\frac{\partial \Phi\left(r,h\right)}{\partial h}$ is the derivative of the potential with respect to the smoothing length. Expressions for the softened potential and its derivatives are tabulated in (\cite{PriceMon}).

\section{Initial conditions}

\begin{table*}
\centering
 \begin{minipage}{140mm}
  \caption{Summary of the physical parameters and the final outcome for the models considered in this paper. The initial mass, radius and density for each model are given by the constant values 5 x 10\superscript{16} cm, 1 $M_{\bigodot}$ and 3.8 x 10\superscript{-18} g/cm\superscript{3},respectively. }
 \begin{center}
  \begin{tabular}{cccccc}
  \hline
Model & Temperature(K) & $c_{0}$(cm/s) & $\rho_{crit}$(g/cm\superscript{3}) & Final outcome & Binary separation(AU) \\
A  & 8  & 1480.0 & 5 x 10\superscript{-14} & Binary & 278.685 \\
B  & 9  & 1570.0 & 5 x 10\superscript{-14} & Binary & 343.235 \\
C  & 10 & 1650.0 & 5 x 10\superscript{-14} & Binary & 378.364 \\
D  & 11 & 1730.0 & 5 x 10\superscript{-14} & Single & --- \\
E  & 12 & 1810.0 & 5 x 10\superscript{-14} & Single & --- \\
F  & 10 & 1650.0 & 5 x 10\superscript{-15} & None   & --- \\
G  & 10 & 1650.0 & 5 x 10\superscript{-13} & Triple & --- \\
H  & 11 & 1730.0 & 5 x 10\superscript{-14} & Binary & 228.438 \\
I  & 12 & 1810.0 & 5 x 10\superscript{-14} & Binary & 197.608 \\ 
\hline 
\end{tabular}
\end{center}
\end{minipage}
\end{table*}

The initial conditions for the cloud collapse calculations considered in this a paper are variants of the Boss and Bodenheimer collapse test with initial conditions described in (\cite{Burkert}). We used this test previously to validate the GRADSPH code on collapse calculations(see section 8.2.3 in \cite{GRADSPH}). 
The initial condition that we take as our standard model is a solar mass cloud with uniform density and radius R = 5 x 10\superscript{16} cm.  The cloud is assumed to be in solid body rotation around the z-axis of the coordinate system with an angular velocity equal to $\omega$ = 7.2 x 10\superscript{-13} rad/s and is rotating counter clockwise. The mean initial density of the cloud is
$\rho_{0}$ = 3.8 x 10\superscript{-18} g/cm\superscript{3}. The chemical composition of the cloud is assumed to be a mixture of hydrogen and helium gas with mean molecular weight $\mu$ = 3. The initial condition is characterized by the parameters $\alpha$ and $\beta$, which correspond to the ratio of thermal and kinetic energy with respect to the gravitational potential energy of the cloud.  These parameters are defined as follows:

\begin{equation} \label{alpha}
\alpha=\frac{5 R k T}{2 G M \mu m_{h}},
\end{equation}

\begin{equation} \label{beta}
\beta=\frac{R^{3}\omega^{2}}{3 G M},
\end{equation} 

where $G$ is the gravitational constant, $k$ is the Boltzmann constant, and $m_{h}$ denotes the mass of the hydrogen atom. For the standard initial temperature of 10 K, the initial values 
for $\alpha$ and $\beta$ are 0.26 and 0.16, respectively. In our models, the value of $\alpha$ changes by modifying the initial temperature, whereas $\beta$ will be kept fixed at its standard value. The mean free-fall time of the initial condition is given by 

\begin{equation} \label{freefalltime}
t_{ff}=\sqrt{\frac{3 \pi}{32 G \rho_{0}}}
\end{equation}

and is 33968 years for the standard initial condition defined above. The initial setup is implemented in our SPH code by placing equal-mass particles on a hexagonal closely packed lattice and retaining only the particles within the initial cloud radius. The code uses internal dimensionless units which are defined by setting G=M=R=1.
To initiate fragmentation of the cloud models, we add an azimuthal density perturbation to the uniform initial condition with a certain mode number m and amplitude A which has the following form:

\begin{equation} \label{densityperturbation}
\rho=\rho_{0}\left(1+A sin\left(m \phi \right)\right],
\end{equation}

where $\phi$ is the azimuthal angle in spheriçal coordinates $\left(r,\phi,z\right)$. We implement this perturbation by perturbing the azimuthal angle of the SPH particles to a new angle $\phi^{*}$ which is determined by solving the equation

\begin{equation} \label{phiperturbed}
\phi=\phi^{*}+\frac{A sin\left(m \phi^{*}\right)}{m}.
\end{equation}

A common mode number m = 2 is adopted for all the models listed in table I while azimuthal density perturbations of amplitude A = 0.1 and A = 0.25 are considered for models (A, B, C, D, E, F, G) and models (H and I), respectively. The thermodynamic behaviour of the gas during cloud collapse is approximated by adopting a barotropic equation of the state as suggested by (\cite{Tohline}) and (\cite{Matsunaga}).
The pressure and sound velocity of the gas are given by the expressions

\begin{equation} \label{Pressure}
P=\rho c_{0}^{2}\left[1+\left(\frac{\rho}{\rho_{crit}}\right)^{\gamma-1}\right],
\end{equation}

and 

\begin{equation} \label{soundvelocity}
c_{s}=c_{0}\left[1+\left(\frac{\rho}{\rho_{crit}}\right)^{\gamma-1}\right]^{1/2},
\end{equation}

where $c_{0}$ is the initial sound velocity, $\rho_{crit}$ is the critical density which determines the transition point from isothermal to adiabatic behaviour of the gas, and
$\gamma$ is the adiabatic exponent which we set to 5/3. Our standard value for the critical density is $\rho_{crit}$ = 5 x 10\superscript{-14} g/cm\superscript{3}.
The initial models are allowed to evolve under the action of the short-range hydrodynamical forces and the self-gravity of the cloud. No external radiation feedback is taken into consideration during evolution of our models.

In order to avoid artificial fragmentation in our models because of insufficient resolution, we need to satisfy the resolution criterion derived by (\cite{BateBurkert}) in our models. This is achieved by keeping the minimum resolvable mass of the code, which equals $M_{min}=2 N_{opt}m $, below the local Jeans mass.  Since the Jeans mass decreases during the isothermal collapse stage and increasing again when the gas becomes optically thick and adiabatic, this condition defines an upper limit to the mass of an SPH particle following (\cite{Arreaga}) and (\cite{GRADSPH}):

\begin{equation} \label{particlemass}
m < \frac{\pi^{3/2} c_{0}^{3}}{2 N_{nopt}\rho_{crit}^{1/2} G^{3/2}}.
\end{equation}

The present numerical study is conducted by taking a total number of SPH particles equal to N=250025 in the initial condition and setting the number of neighbours to $N_{opt}$=50.
Table I provides an overview of the models which have been used to examine the evolution of the binary separation as a function of the initial temperature of the clouds as well as the impact of the changes in the critical density along with changes in perturbation amplitude. The table gives, for the 9 models labeled A-I, the values of the initial temperature, the sound velocity, the critical density, the resulting binary separation when applicable, as well as the final outcome of the simulations which will be discussed in section 4. It can be seen that models A-E explore the effect of changes in the temperature with the critical density kept at its standard value. In models F and G, on the other hand, the initial temperature is set to 10 K and $\rho_{crit}$ takes on different values to examine the effect of this parameter on the evolution of the cloud models. The last two models labeled H and I reveal the effects of strength of perturbation amplitude on evolution of the collapsing cloud.

In our models, fragmentation proceeds in two stages. In the first stage, a rotating disk like structure forms in the center of the cloud as a result of the conservation of angular momentum. In the second stage, this disk fragments into protostellar fragments once the gas has become optically thick and evolves adiabatically because of non-axisymmetric gravitational instabilities. The growth of this kind of gravitational instabilities is governed by the Toomre parameter for a disk with surface density $\Sigma$ and epicyclic frequency $\kappa$(\cite{Toomre}). This parameter is defined as 

\begin{equation} \label{ToomreQ}
Q=\frac{c_{s}\kappa}{\pi G \Sigma},
\end{equation}

and the value of Q must be smaller than unity to enable non-axisymmetric instabilities to grow within the disk. Plots of the radial behaviour of the Toomre parameter at various stages of the collapse will be discussed in section 4. For visualization of the results of our simulations, we use the visualization tool SPLASH developed and made publicly available to the community by Daniel Price(\cite{SPLASH}).

\section{Results and discussion}

\begin{table*}
\centering
\begin{minipage}{140mm}
 \begin{center}
 \begin{tabular}{ccccc}
  \hline
 Model & A & B & C & D \\ 
 $t_{max}$(yrs) & 47913.746 & 47913.746 & 47913.746 & 47913.746 \\
 $\rho_{max}$(g/cm\superscript{3}) & 4.767 x 10\superscript{-11} & 2.380 x  10\superscript{-11} & 1.883x10\superscript{-11} & 7.551x10\superscript{-12} \\
 $T_{max}$(K) & 1298 & 922.5 & 878.4 & 530.2 \\
\hline
\end{tabular}
\begin{tabular}{cccccc}
  \hline
 Model & E & F & G & H & I \\ 
 $t_{max}$(yrs) & 47913.746 & 47913.746 & 41443.078 & 47913.746 & 47913.746 \\
 $\rho_{max}$(g/cm\superscript{3}) &  4.643x10\superscript{-12} & 8.118x10\superscript{-14} & 8.909 x 10\superscript{-10} & 7.787 x 10\superscript{-12} & 5.042 x 10\superscript{-12}\\
 $T_{max}$(K) & 421.5 & 116.7 & 2456 & 540.965 & 444.7 \\
\hline
\end{tabular}
\caption{Summary of the maximum evolution time, as well as the maximum density and the maximum temperature obtained from the evolution of the models. }
\end{center}
\end{minipage}
\end{table*}

Figure 1 shows successive column density maps of the evolution of models A-E at 5 different times indicated to the left of the columns.  The initial temperature of the models is indicated at the top of each column. The column density maps are top down views in the xy plane where the density distribution has been integrated along the rotational axis of the cloud.  Figure 2 shows the corresponding edge-on column density maps in the xz plane orthogonal to the rotational axis. Every single square panel in figure 1 has a physical dimension of approximately 401 x 401 AU. In Figure 2 every single panel has physical dimensions of approximately 401 x 167 AU, respectively. The outcome of each simulation model and the separation of the binary at the end of each run are summarized in table I. Table II contains the maximum time, $t_{max}$, expressed in years, during which we have been able to evolve the model before the timestep became prohibitively small, as well as the maximum density $\rho_{max}$  and the maximum temperature $T_{max}$ attained during the evolution of each model.
The rise in initial temperature going from model A to model E has a drastic impact on the evolution of the cloud collapse.  Models with initial temperatures up to standard value of 10 K develop binary systems, whereas molecular cores with temperatures above 10 K fail to develop binaries yielding only single protostars instead. Changing the initial temperature also has a big impact on the separation of the binaries in the range $8K < T <= 10 K$. For example, raising the temperature from  8 to 9 K leads to change in the final binary separation of around $\Delta d$ = 64 AU, while a further increase from 9 to 10 K leads to a change around $\Delta d$ = 35 AU. The panels in Figure 2 also clearly show the formation of the intermediate disk structure which fragments into protostars for $T <=10 K$.

The phase in which the collapsing fragments are still connected by a bar-like structure is comparatively longer in molecular cloud cores with temperatures below the conventional value of 10 K. This is illustrated in Figure 1 by the snapshots  a5, b5, and c5. It can be seen that this connecting filament starts to disappear earlier in model C than in models A and B.
There is also an important difference in evolution of the orbital elements of the models with increasing temperature for models A,B and C which produce binary systems as the final products.
The left panel of Figure 3 shows the evolution of the binary separation with time for these models. The time is indicated in units of the initial freefall time of the cloud. Each model shows a qualitatively similar evolution showing a minimum in the binary separation, after which the separation increases up to the end of the simulation. The physical reason for this behavior is likely connected to the accretion of high angular momentum gas by the protostars from the surrounding envelope. However, there are important differences in terms of the speed of this evolution and the minimum separation of the binaries. Colder molecular cloud evolve slower than hotter ones in the sense that the minimum separation is attained later in the course of the evolution. Note that at the end of the evolution, the faster evolving hotter cores have reached larger binary separations than the colder ones. 

We approximately estimate the masses of the evolving fragments by locating the particle with the highest density $\rho_{max}$ within each fragment and include all particles whose density is higher than $f \rho_{max}$, where $f$ is set to 0.001. The velocity and position of each fragment is determined from the position and velocity of the center of mass of the clump, respectively. A similar procedure has been adopted by \cite{Arreaga}. Finally, the orbital elements of the binaries are determined from the positions, velocities and masses of the fragments.

Figure 3 shows that the binary separation in models (A, B, C) mainly follows the variations in semi-major axis and is less affected by the evolution of the orbital eccentricities which remain high ($\sim 0.5-0.6$) throughout of the evolution. Figure 4, on the other hand, shows that the initial thermal state of the clouds in decisive in determining the final mass fraction included in the binary system.  The cooler the core, the more mass is involved in the binary fragmentation. The coolest model A has more than 
50 $\%$ of its mass included at t=1.35 $t_{ff}$. Since about half of the mass of the cloud is at most included in the binaries at the end of our simulations, we expect that the orbital elements of the binaries reported after 1.4 $t_{ff}$ are unlikely to represent the final orbital elements but are rather indicative of the trend of their further evolution. In particular, we expect that the increasing trend of the binary separation will continue despite the fact that we were unable to prolong the simulations for much longer than 1.4 $t_{ff}$. We also note that the decline of the mass fraction which can be seen in the lower left panel in Figure 4, but is only temporary for model C because this model evolves faster, is likely caused by mass exchange between the fragments and their surrounding disk like structure which could be the result of the tidal forces exerted by the companions. The symmetry of the initial conditions ($m=2$) implies that the binary mass ratio is always close to unity. 

Figure 5 shows the time evolution of the maximum temperature for models A-E (left panel) as well as the time evolution of the maximum of the density in the right panel. Comparison of the models shows that for colder molecular cloud cores, the clouds are subject to an earlier increase in temperature along with a faster transition to the adiabatic regime. It is also evident that the final temperature of the models is highly dependent on their initial thermal state with colder models reaching higher final temperatures.
The logarithmic density profiles shown in the right panel indicate a similar trend for the maximum density. Here, the colder models reach higher final densities because self-gravitating fragments which have reached the adiabatic heating phase are formed earlier during the course of their evolution.

Figures 6 to 8 respectively show top down views of the evolution of models H and I as well as the evolution of their maximum density and temperature, respectively. In these models, the temperature is raised above the nominal value of 10 K( 11 K for model H and 12 K for model I) and a bigger perturbation amplitude is applied(A=$25 \%$). Because of the increased strength of the perturbation, models H and I form binary systems at the end of the simulation, and thus clearly indicate that the fragmentation process is a function of both the thermal energy of the core and the strength of the initial perturbation. 

Comparison of the results of models A-C and H and I leads to several conclusions. Firstly, cold molecular cores ($T_{0}=$8 K, 9 K, 10 K) with a smaller amplitude of the initial azimuthal density perturbation (A = 10 $\%$) give birth to protobinary systems that evolve relatively quickly and include protostellar fragments that reach higher densities and temperatures compared to those in models H and I. The binaries in models A-C attain minimum separations of $\sim 100$ A.U. and then increase in separation due to the accretion of high angular momentum material.
However, upon examining the evolution of the relatively warm molecular cores in models H and I(11 K and 12 K, respectively) with a larger amplitude of the initial azimuthal density perturbation(A=25 \%), we find that the overall speed of the evolution of these fragmenting models is much slower than for models A-C. At the end of the simulation, models H and I have not yet experienced pericenter passage, and the resulting densities and temperatures are lower than for models A-C, as can be seen by comparing the left and right panels of Figures 5 and 8, respectively. Over the course of the simulation, models H and I undergo a monotonic decline in excentricity from $\sim 0.7$ to $\sim 0.5$.

\begin{figure*}
\centering
% \begin{minipage}{200mm}
\includegraphics[trim = 0mm 0mm 0mm 0mm, clip, width=5in]{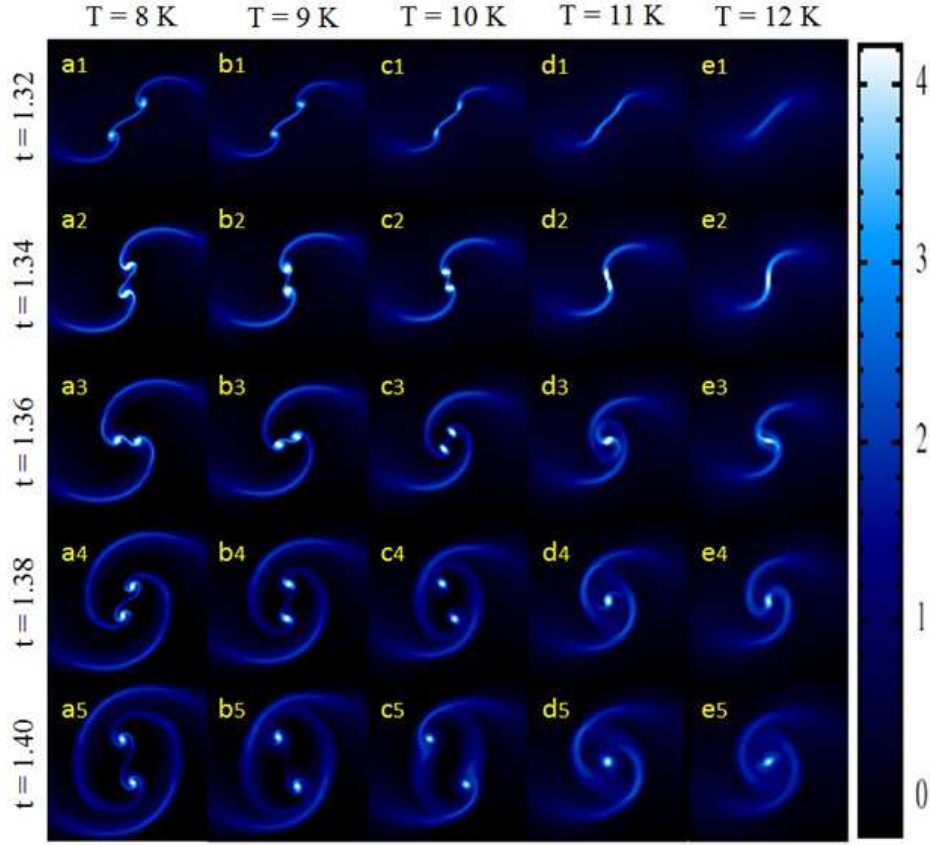}
\caption{Simulation results for models A, B, C, D and E. Each column of plots shows face-on views of the column density integrated along the z axis for each particular model. The corresponding initial temperature is shown at the top of each column. The times corresponding to each row are given in units of $t_{ff}$. The binary separations for each snapshot are respectively given by (a1) d= 313.184 AU, (a2) d = 224.943 AU,(a3) d = 88.758 AU, (a4) d = 199.348 AU, (a5) d = 278.685 AU, 
(b1) d = 282.382 AU, (b2) d = 200.013 AU,(b3) d = 151.047 AU, (b4) d = 251.859 AU, (b5) d = 343.235 AU, (c1) d = 262.515 AU, 
(c2) d = 166.739 AU,(c3) d = 173.062 AU, (c4) d = 266.667 AU, (c5) d = 378.364 AU. The horizontal and vertical
dimensions of each plot in the xy-plane are 0.12 x 0.12 in dimensionless units. The color bar on the right shows $log\left(\Sigma \right)$ in 
dimensionless units. Each calculation was performed with 250025 SPH particles. }
\label{fig1}
% \end{minipage}
\end{figure*}

\begin{figure*}
% \begin{minipage}{200mm}
\centering
\includegraphics[trim = 0mm 20mm 0mm 0mm, clip, width=5in]{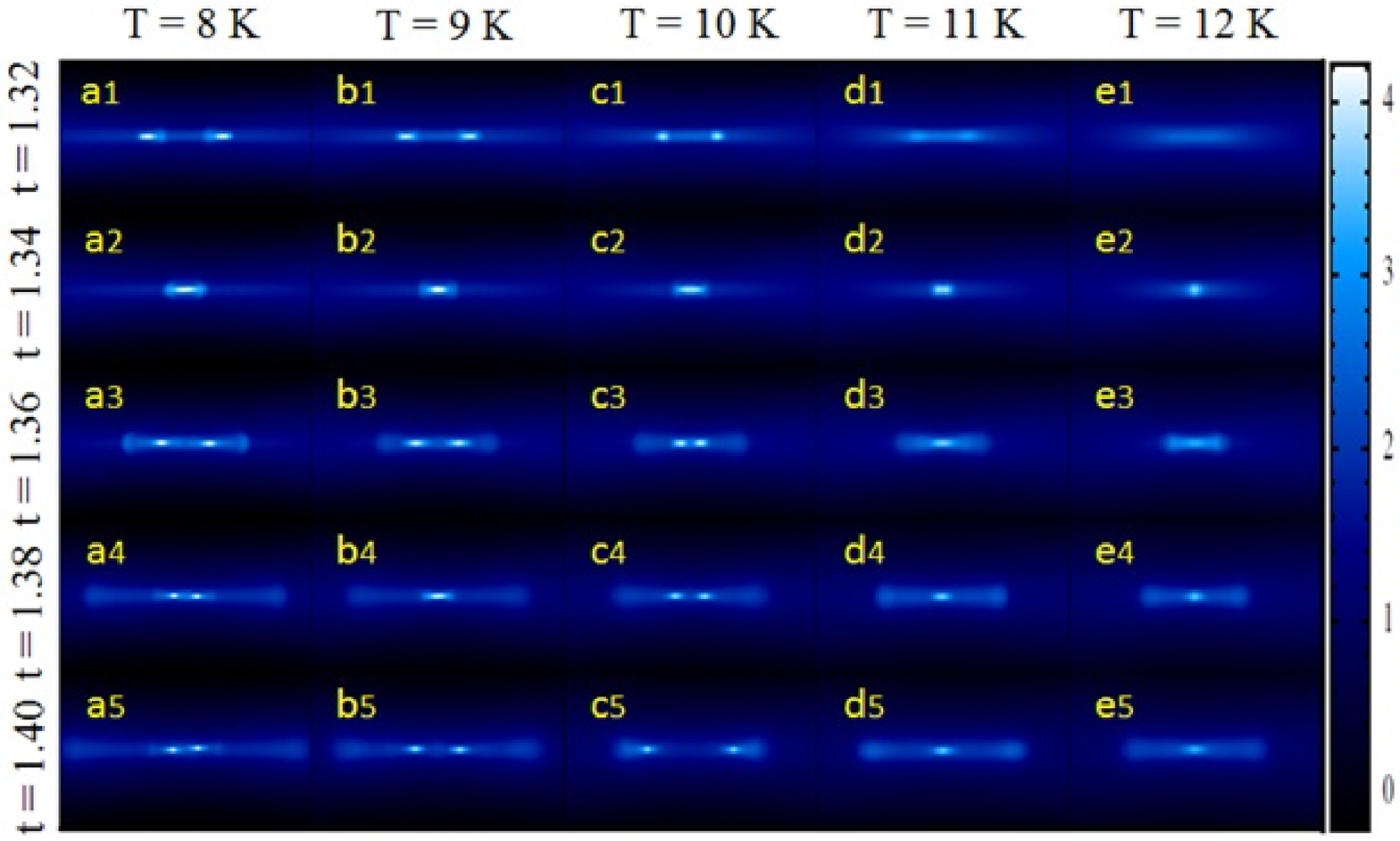}
\caption{Same as in Figure 1, but here we show edge-on views of the column densities integrated along the direction orthogonal to the z-axis. The horizontal and vertical dimensions in the xz-plane in each panel are 0.1 x 0.05 in dimensionless units, respectively. }
\label{fig2}
% \end{minipage}
\end{figure*}

Let us now take a closer look at the evolution of the protostellar disks in models A,B and C. The time evolution of the Toomre parameter is an important indicator of the growth of gravitational 
instabilities in the disks which form during the gravitational collapse of our cloud models with values below 1 indicating the onset of gravitational instabilities. 
Figures 9a to 9e contain plots of the radial evolution of the Toomre parameter, averaged with respect to the azimuthal angle in the midplane of the disk, for models A-E at various instants of time.
It can be observed that the time evolution of the Toomre parameter closely follows the development of fragments during the collapse of the cloud models. At the earlier stages of the evolution, all models exhibit gravitational instability within a radial range of 167 AU.
As the clouds evolve further, the gravitational instabilities in models A,B and C, which start their evolution from temperatures at or below 10 K, gradually sweep larger parts of the disks, almost doubling their active radial range up to 334 AU. On the other hand, The hotter models D and E do not show this kind of behaviour. 
For the relatively cold gas models, the part of the initial molecular core which is subject to gravitational fragmentation is about 10 percent of the initial radius. Beyond this range the disk seems fairly stable and remains free of fragmentation processes. 
The time evolution of Q for each model also shows that molecular cloud cores with higher initial temperatures take more time to undergo fragmentation than models which start their evolution with lower initial temperature values. The threshold value of unity at or below which fragmentation should happen
never seems to be reached outside of the inner regions by models D and E, whereas models A to C, which start their evolution from temperatures ranging from 8 K to 10 K, show a drop of the Q values well below unity and hence give rise to self-gravitating fragments which are well under way to reach protostellar densities.

\begin{figure*}
% \begin{minipage}{200mm}
\centering
\includegraphics[trim = 0mm 0mm 0mm 0mm, clip, width=5in]{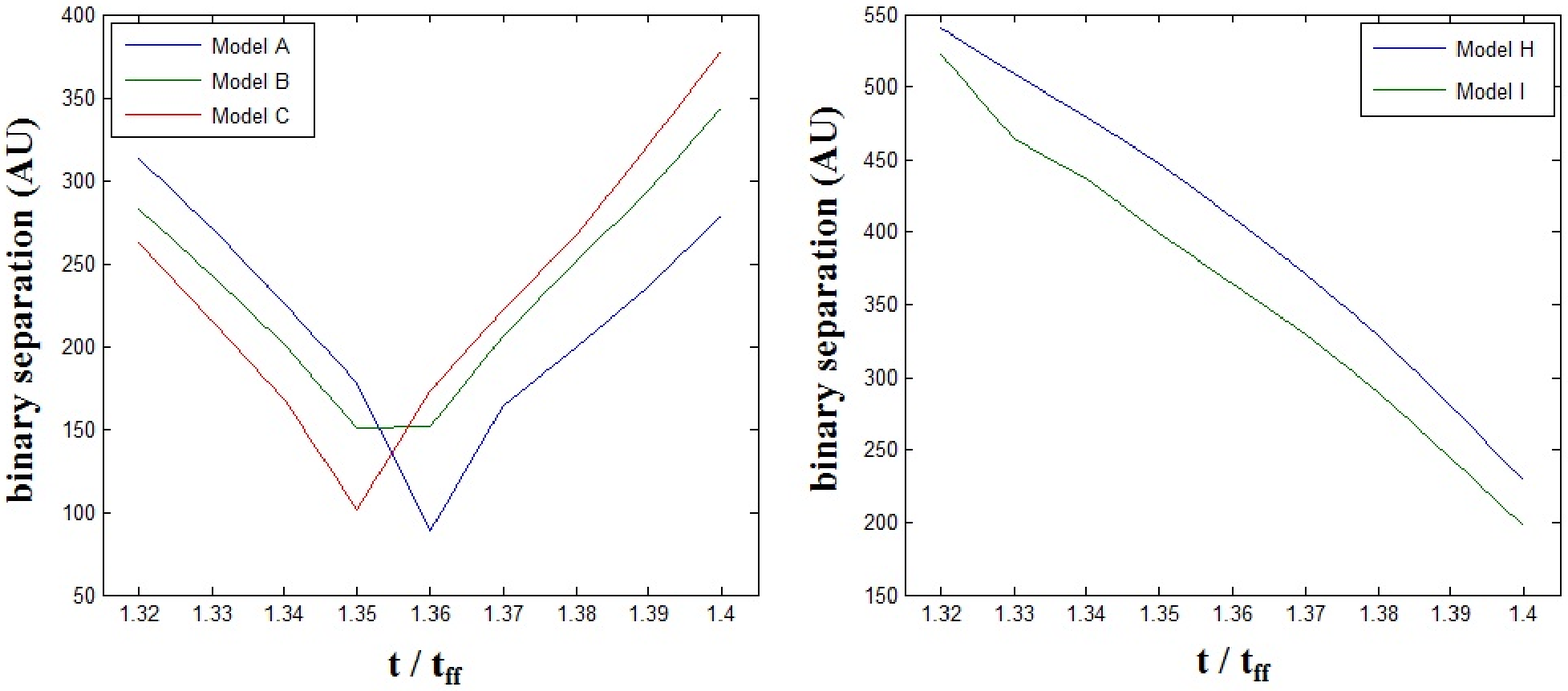}
\caption{Evolution of the binary separation for models A, B, and C(left panel) and for models H and I(right panel)
as a function of time up to 1.4 $t_{ff}$. The binary separation for the two 
protostars is determined by searching for the SPH particles with maximum density and their corresponding distance in simulation space. }
\label{fig3}
% \end {minipage}
\end{figure*}

\begin{figure*}
% \begin{minipage}{200mm}
\centering
\includegraphics[trim = 0mm 0mm 0mm 0mm, clip, width=3in]{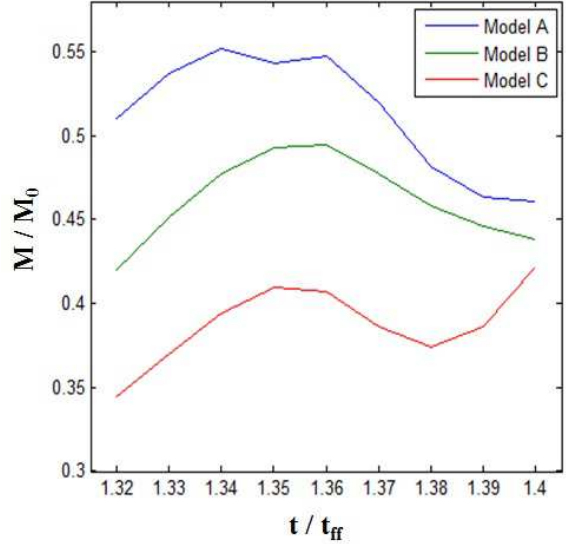}
\caption{Evolution of the total relative mass fraction for the protostellar binaries in 
models A, B, and C as a function of time up to 1.4 $t_{ff}$. }
\label{fig4}
% \end {minipage}
\end{figure*}

\begin{figure*}
% \begin{minipage}{200mm}
\centering
\includegraphics[trim = 0mm 0mm 0mm 0mm, clip, width=5in]{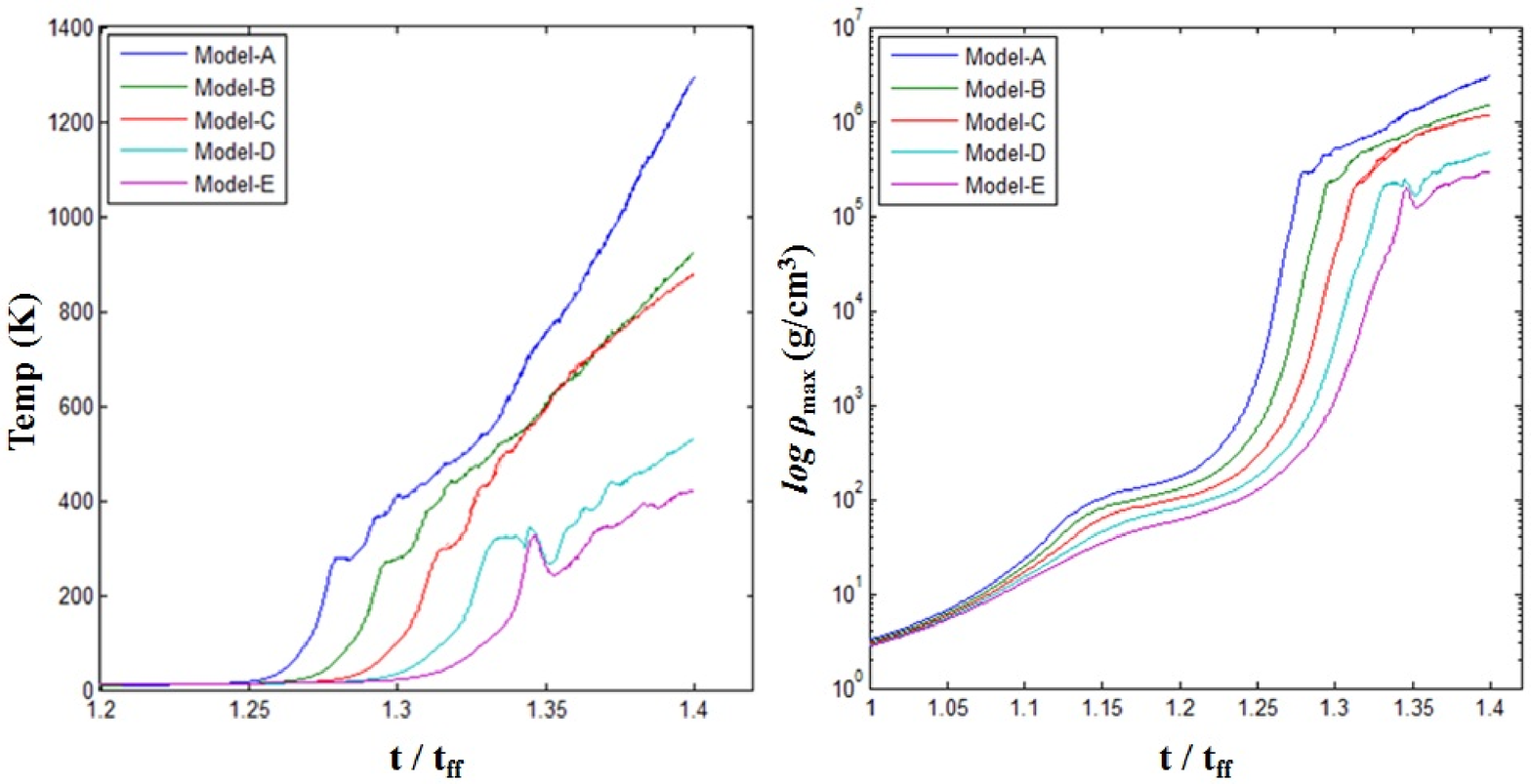}
\caption{Time evolution of the maximum temperature within  
the collapsing cloud for models A, B, C, D, E (left panel),and the time evolution 
of the maximum density for the same set of models (right panel).} 
\label{fig5}
% \end{minipage}
\end{figure*}

\begin{figure*}
\centering
% \begin{minipage}{200mm}
\includegraphics[trim = 0mm 0mm 0mm 0mm, clip, width=5in]{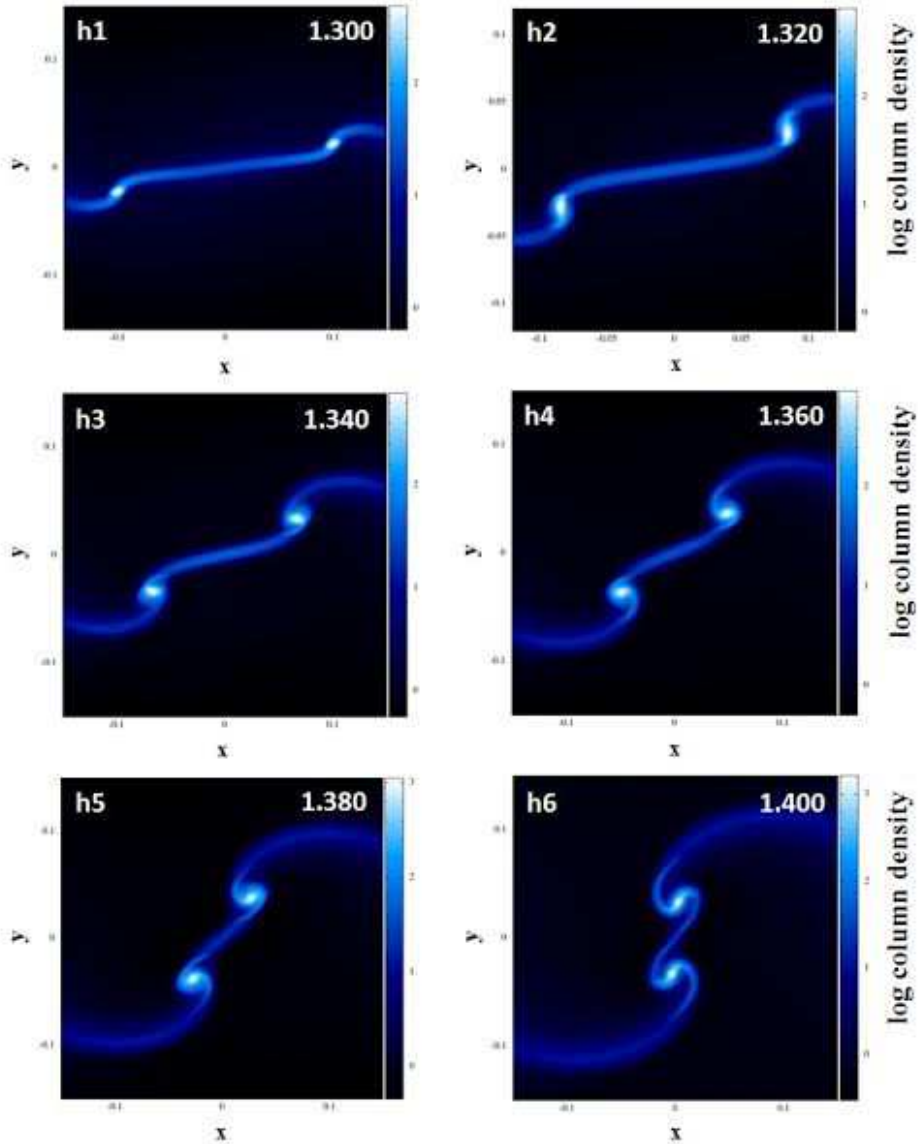}
\caption{Simulation results for model H. Each plot shows a face-on view the column density integrated along the z axis. The times corresponding to each row are given in units of $t_{ff}$. The binary separations for each snapshot are respectively given by (h1) d= 590.130 AU, (h2) d = 540.667 AU,(h3) d = 479.196 AU, (h4) d = 410.897 AU, (h5) d = 328.628  AU, (h6)= 228.438 AU. The horizontal and vertical dimensions of each plot in the xy-plane are 0.15 x 0.15 in dimensionless units. The color bar on the right shows $log\left(\Sigma \right)$ in dimensionless units. Each calculation was performed with 250025 SPH particles. }

\label{fig6}
% \end{minipage}
\end{figure*}

\begin{figure*}
\centering
% \begin{minipage}{200mm}
\includegraphics[trim = 0mm 0mm 0mm 0mm, clip, width=5in]{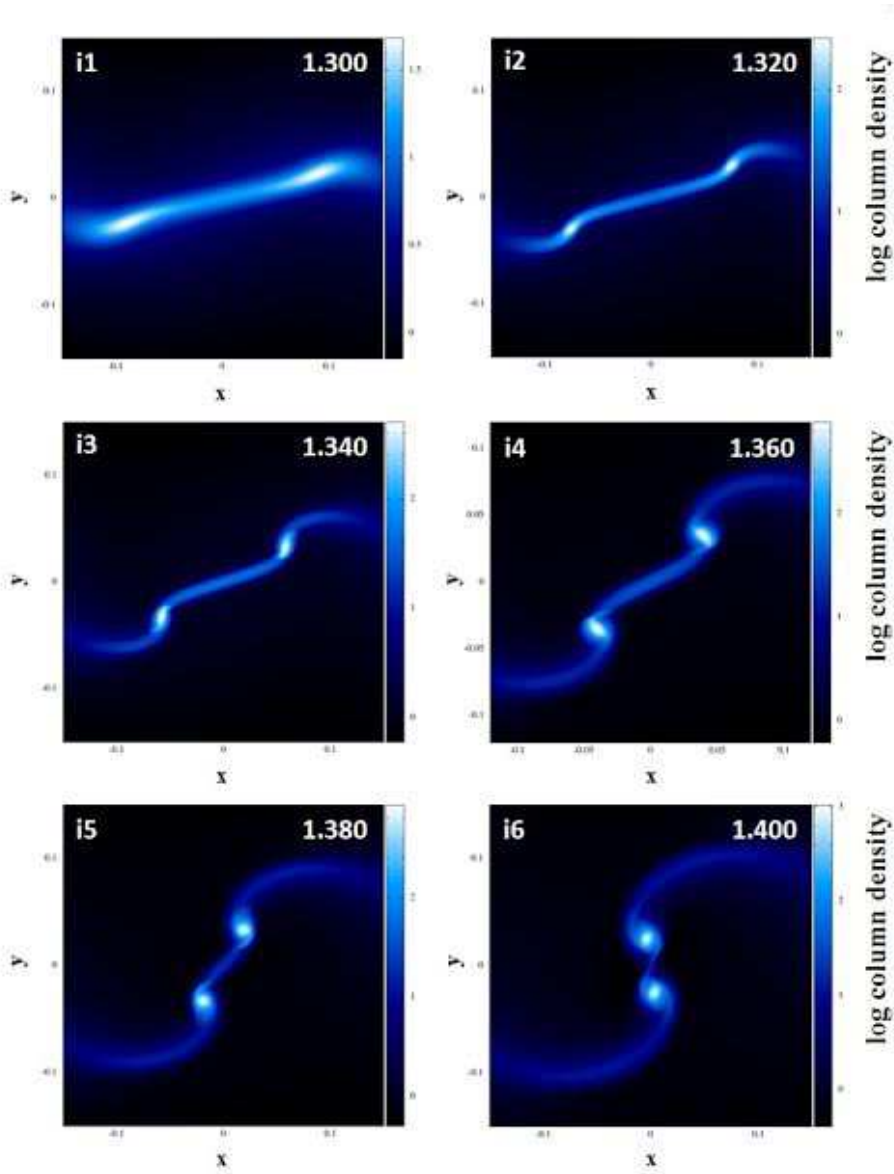}
\caption{Simulation results for model I. Each plot shows a face-on view the column density integrated along the z axis. The times corresponding to each row are given in units of $t_{ff}$. The binary separations for each snapshot are respectively given by (i1) d= 531.065 AU, (i2) d = 522.757 AU,(i3) d = 435.9223 AU, (i4) d = 364.861 AU, (i5) d = 289.241 AU, (i6)= 197.608 AU. The horizontal and vertical dimensions of each plot in the xy-plane are 0.15 x 0.15 in dimensionless units. The color bar on the right shows $log\left(\Sigma \right)$ in dimensionless units. Each calculation was performed with 250025 SPH particles. }

\label{fig7}
% \end{minipage}
\end{figure*}

\begin{figure*}
% \begin{minipage}{200mm}
\centering
\includegraphics[trim = 0mm 0mm 0mm 0mm, clip, width=5in]{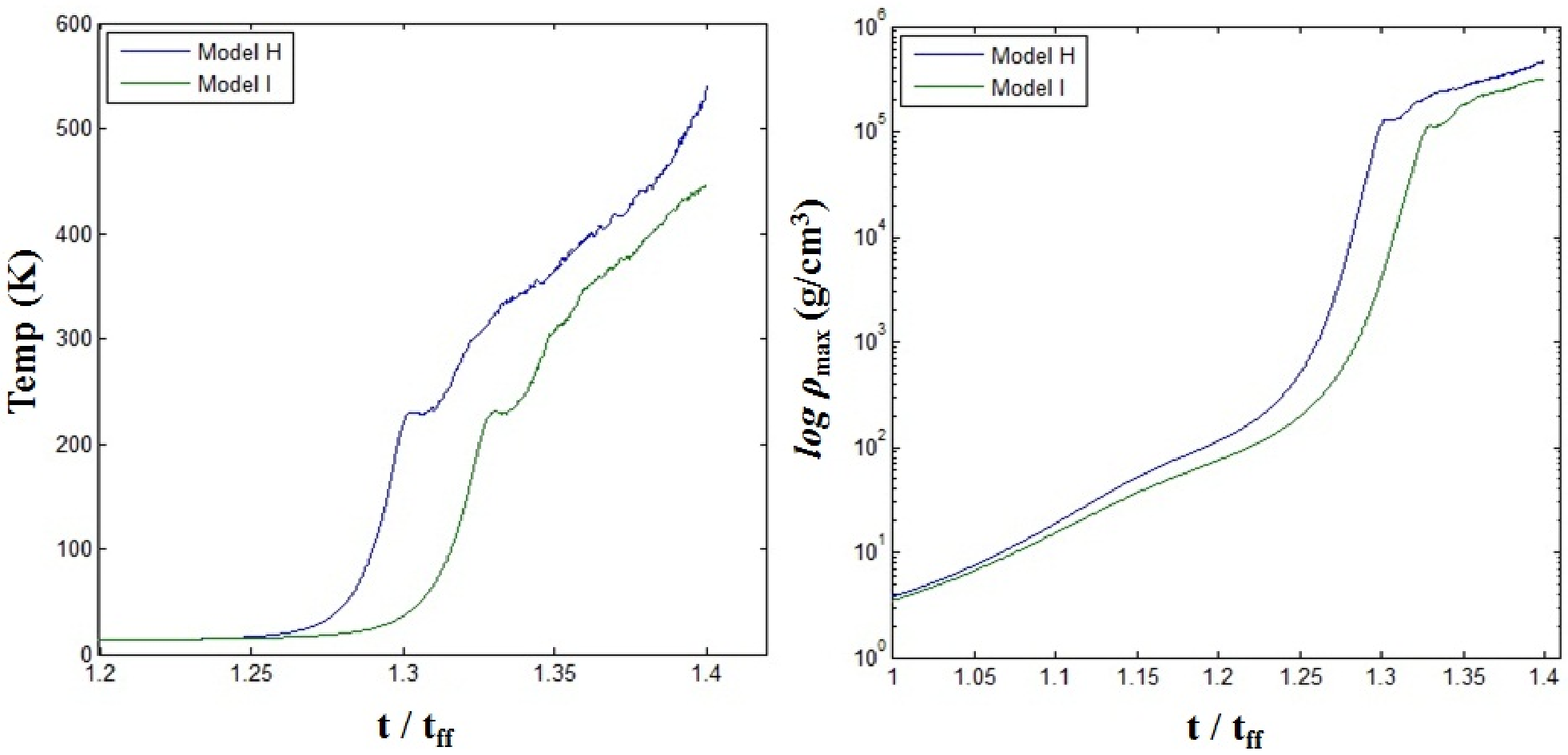}
\caption{Time evolution of the maximum temperature within  
the collapsing cloud for models H and I(left panel),and the time evolution 
of the maximum density for the same set of models (right panel). }
\label{fig8}
% \end {minipage}
\end{figure*}

\begin{figure*}
% \begin{minipage}{200mm}
\centering
\includegraphics[trim = 0mm 0mm 0mm 0mm, clip, width=4.0in]{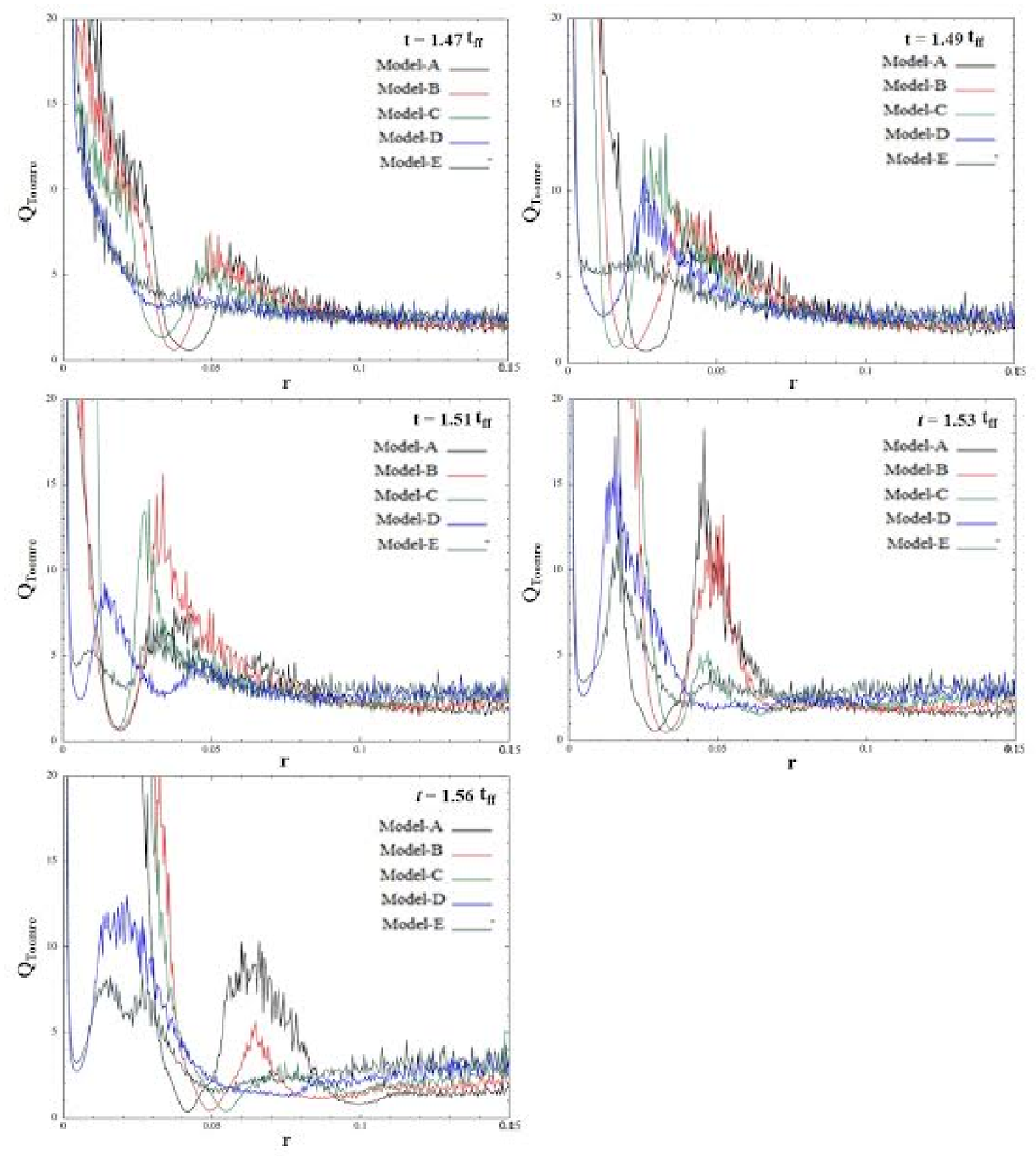}
\caption{Traces of the azimuthal average of the Toomre parameter values as a function of the radial distance r in the cloud midplane for models A to E at times $t=1.47 \, t_{ff}, t=1.49 \, t_{ff},t=1.51 \, t_{ff}, t=1.53 \, t_{ff}$, and $t=1.56 \, t_{ff}$, respectively. }
\label{fig9}
% \end{minipage}
\end{figure*}

We also made an attempt to understand the sensitivity of the structural evolution of collapsing molecular cores 
on the threshold density that marks the beginning of the adiabatic heating phase. 
At that moment the collapsing core becomes optically thick enough to trap heat inside leading to a transition from isothermal to adiabatic behaviour of the collapsing gas.
For this purpose, we varied the value of the critical density $\rho_{crit}$ by an
order of magnitude around 5 x 10\superscript{-14} g/cm\superscript{3} and investigated the role of adiabatic heating in determining the evolution of the models. The outcome of this investigation is illustrated by the results for models F and G, for which the critical density is set to values 10 times lower or 10 times larger than the standard value, respectively. The evolution for model G is illustrated in Figures 10 and 11. Model G follows a delayed isothermal phase during which a thin but very dense bar is formed connecting the two primary fragments. This bar-like structure subsequently fragments into multiple fragments. This interesting phenomenon also happened in a collapse model test described in (\cite{Kitsionas}), which even further prolonged the isothermal phase by setting $\rho_{crit}$= 5 x 10\superscript{-12} g/cm\superscript{3}.
At the later stages of evolution, the secondary fragments combine with each other and produce a third fragment that starts to interact gravitationally with one of the primary fragments hence giving rise to the possibility of a hierarchical system of protobinaries, with one binary as a component of another already formed binary system.
This final state of the evolution is shown in panel g6 of Figure 10, after which we unfortunately had to stop the calculations because of too small time steps. The emerging protobinary system in model G is found to be the result of a combination of both primary and secondary fragmentation. Note that this evolution is different from that seen in Models A, B, C which resulted in protobinary systems consisting only of primary-primary fragments. For comparison, we refer to panels a5, b5, c5, and g6 of Figure 1 and Figure 10, respectively. Figure 12 shows a detailed view of the protobinary system formed as a result of the secondary fragmentation of the bar. The density evolution of this model shows a quicker shift of eight orders of magnitude in density of the fragments that may evolve up to protostellar densities. This evolution happens in roughly half of the time needed for the rest of the models that yielded a binary system to reach densities near 10\superscript{-10} g/cm\superscript{3}, at which point hydrogen molecules in the collapsing gas experience molecular dissociation and the second collapse phase starts. The sideway views in Figure 13 also clearly show the extent of the vertical dimension of the intermediate disk structure, which is found to be much thinner than for any other collapsing model studied in the present work. Although model G could not evolved to the extent of models of A, B, and C, we suggest that the model is well on its way to becoming a stable triple stellar system.

\newpage

\begin{figure*}
% \begin{minipage}{120mm}
\centering
\includegraphics[trim = 0mm 0mm 0mm 0mm, clip, width=7in]{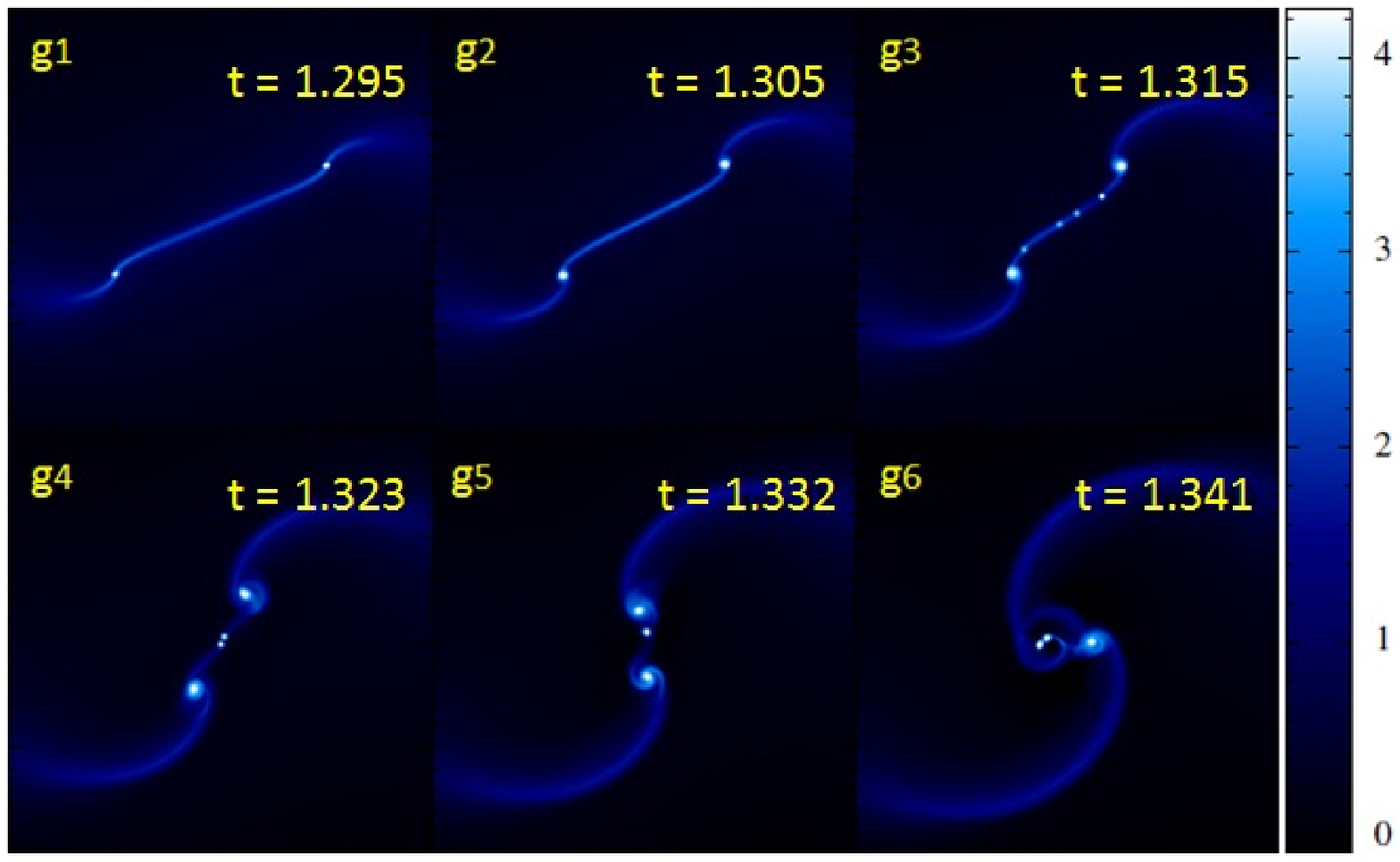}
\caption{Snapshots of column density integrated along the rotational axis during the evolution of model G. The horizontal and vertical dimensions in the xy-plane in each panel are 0.15 x 0.15 in dimensionless units.
The colour bar shows $log\left(\Sigma\right)$ in dimensionless units. 
The times in each row are in units of $t_{ff}$.  }
\label{fig10}
% \end{minipage}
\end{figure*}

\begin{figure*}
% \begin{minipage}{120mm}
\centering
\includegraphics[trim = 0mm 50mm 0mm 0mm, clip, width=7in]{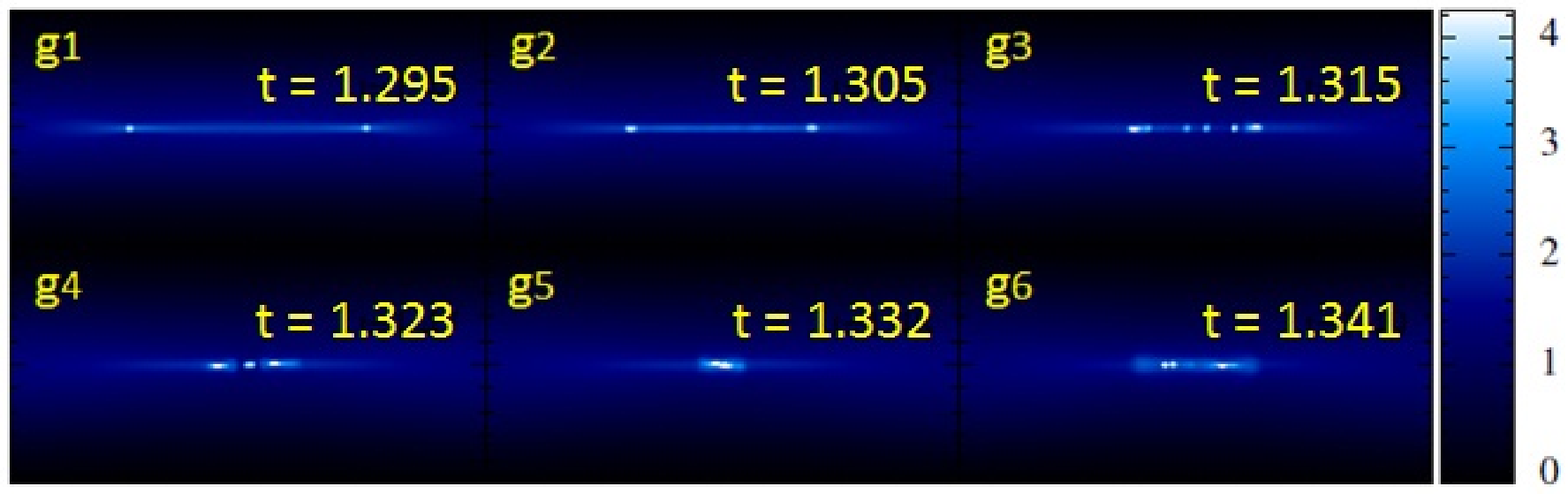}
\caption{Same as in figure 10, but here we show the column densities integrated along the direction orthogonal to the rotational axis.  The horizontal and vertical dimensions in the xz-plane in each panel are 0.1 x 0.05 in dimensionless units. }
\label{fig11}
% \end{minipage}
\end{figure*}

\begin{figure*}
% \begin{minipage}{120mm}
\centering
\includegraphics[trim = 0mm 40mm 0mm 0mm, clip, width=7in]{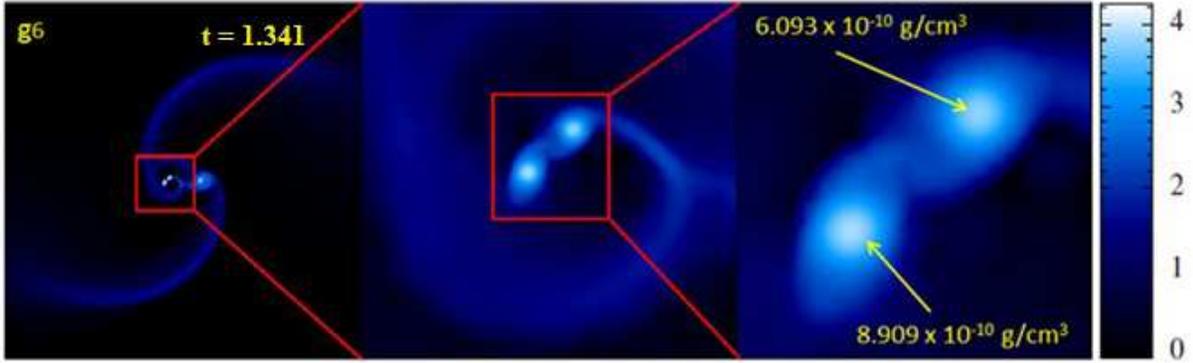}
\caption{Detailed snapshots of 
the column density integrated along the z-axis during the evolution of model G.
The first panel on the left has dimensions 0.25 x 0.25, while the other two panels are successive enlargements of the first one. 
The colour bar on the right shows $log\left(\Sigma \right)$ in dimensionless units. 
The time indicated in each snapshot is given in units of $t_{ff}$. }
\label{fig12}
% \end{minipage}
\end{figure*}

\begin{figure*}
% \begin{minipage}{120mm}
\centering
\includegraphics[trim = 0mm 0mm 0mm 0mm, clip, width=7in]{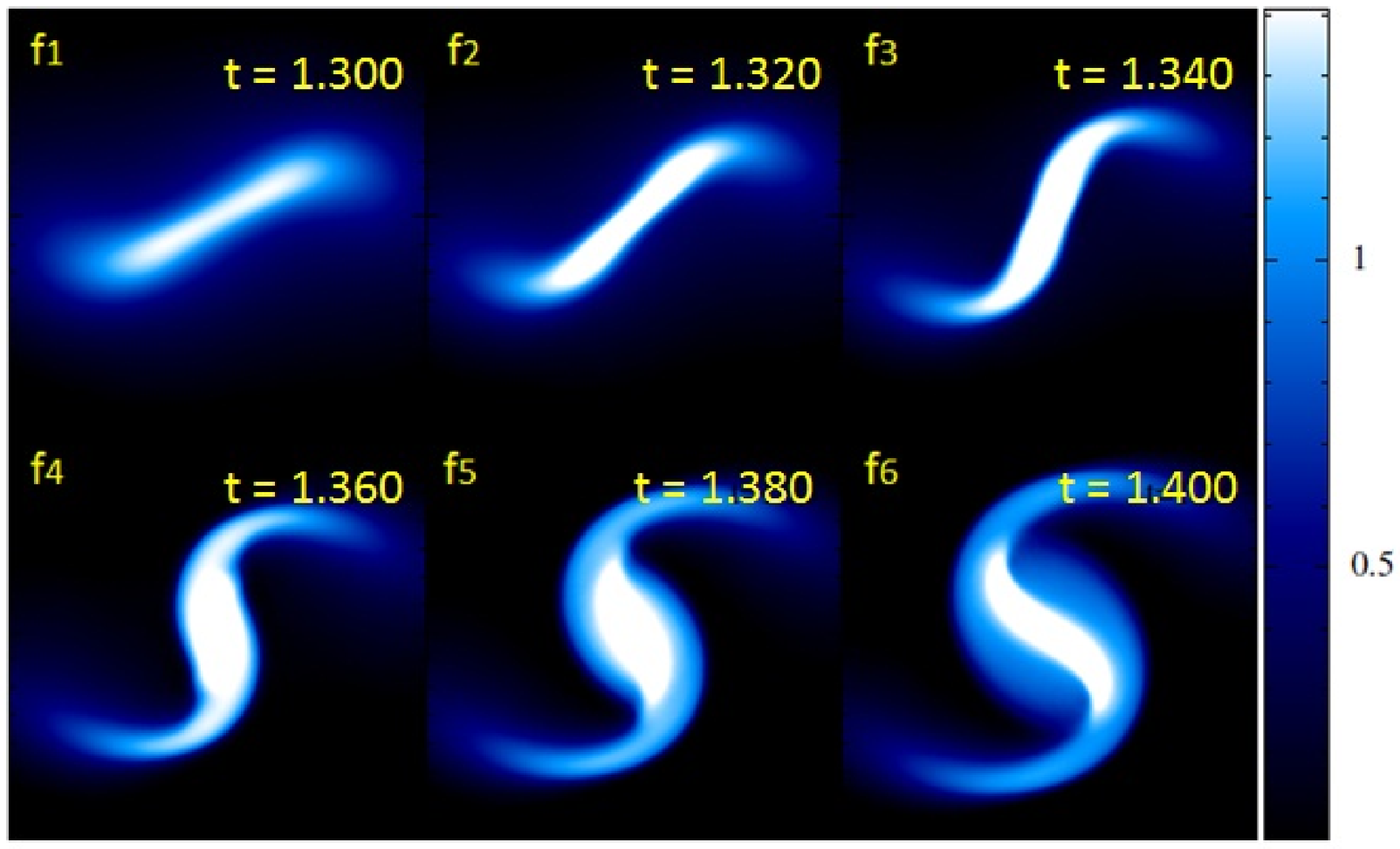}
\caption{Snapshots of column density integrated along the rotational axis during the evolution of model F. The horizontal and vertical dimensions in the xy-plane in each panel are 0.15 x 0.15 in dimensionless units. 
The color bar shows $log\left(\Sigma\right)$ in dimensionless units. 
The times in each row are in units of $t_{ff}$. Each calculation was performed with 250025 SPH particles. } 
\label{fig13}
% \end{minipage}
\end{figure*}

\begin{figure*}
% \begin{minipage}{120mm}
\centering
\includegraphics[trim = 0mm 0mm 0mm 0mm, clip, width=7in]{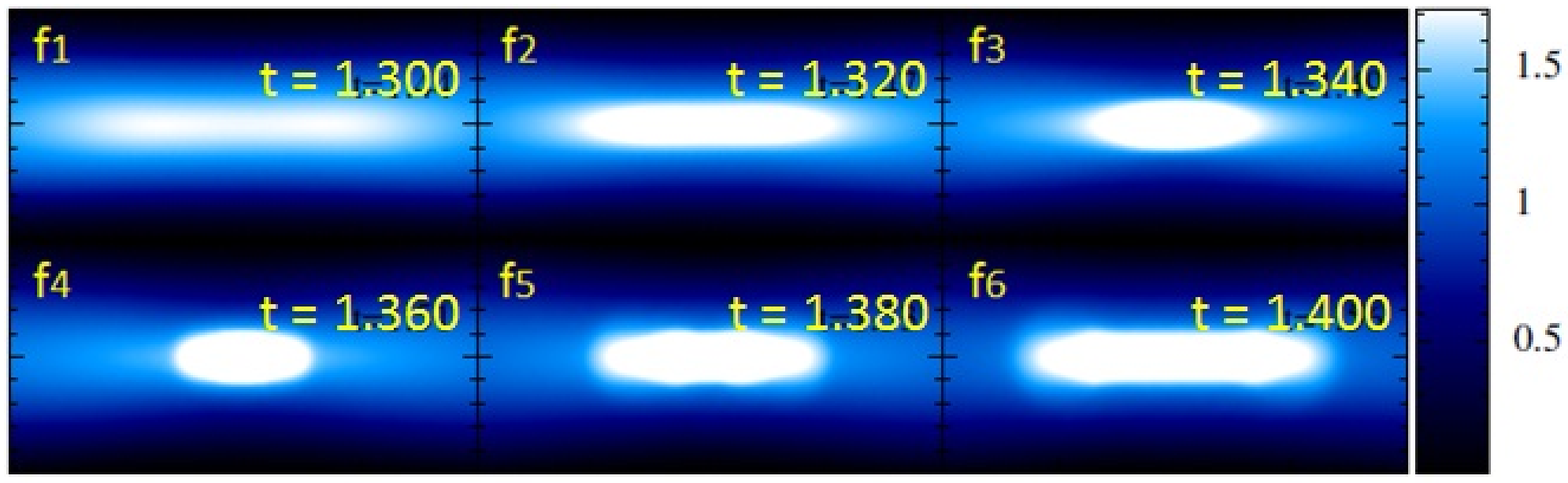}
\caption{Same as in figure 13, but here we show the column densities integrated along the direction orthogonal to the rotational axis. The horizontal and vertical dimensions in the xz-plane in each panel are 0.1 x 0.05 in dimensionless units. }
\label{fig14}
% \end{minipage}
\end{figure*}

Figures 13 and 14 describe the time evolution of model F in the equatorial and vertical planes, respectively. The evolution of the core is free of any process of fragmentation for this model. This mainly happens because the inward gravitational pull is challenged by the quicker rise of the temperature of the core so that the gas can not collapse to the stage where clumps could be formed. Instead model F initially collapses to a bar that keeps on gaining thickness and eventually gives birth to a defused spiral structure that surrounds the dense bar as can be seen in panel f6 of Figure 13. Similarly, we also see in Figure 14 that such a warm molecular core fails to develop a thin disk structure as in model G. 

The evolution of the surface density and Toomre parameter values in models G and F are shown in Figures 15 and 16 as well as Figures 17 and 18, respectively. The surface density evolution seen in models F and G is a clear manifestation of how conditions leading to fragmentation play a key role in transferring angular momentum of the disk outwards resulting in infall of gas that gives rise to increased surface densities of the disk and hence a drop in the Toomre parameter values. This inverse relationship between Toomre parameter value and the surface density is obvious when comparing the corresponding panels in Figures 16 and 17 and Figures 17 and 18.
Values of the Toomre parameter well above unity also suggest that the thick disk resulting from model F never experiences a phase of gravitational instability. 

A comparison of models C, F, G yields further interesting results on the temperature and density
evolution of the molecular core models which are illustrated in Figure 19.
The left and right panels respectively show the evolution of the maximum temperature and density for these three models. It can be 
seen that changing the critical density over an order of magnitude range results in a direct relation  
between the critical density and the resulting maximum temperature and density values reached during the evolution with higher critical densities corresponding to higher temperature and density values. This
is explained by the fact that a little longer isothermal phase during the collapse allows the molecular gas to get compressed to a denser state, which initially gives rise to a thin filament structure along with multiple fragments which later on in the adiabatic phase become full-fledged protostellar objects. On 
the other hand, if the adiabatic phase takes over the initial isothermal regime a little earlier, 
the core might end up as a mere reservoir of gas that never reaches densities corresponding to protostellar systems. 

Finally, Figure 20 shows snapshots of the temperature integrated along the rotational 
axis for the 6 models A-C(the first row) and for models D,E and G(second row) at the final state of evolution of the models. Hardly any temperature difference is found for the two fragments reaching the state of protostars in models A, B, and C.
The single protostellar objects arising in models D and E correspond to the slightly hotter regions in the two leftmost columns in the second row.  However, Model G (the bottom right panel) represents a significant 
difference in temperature of the resulting triple system mainly because the system hosts both primary and
secondary fragments which have delayed isothermal phases. Moreover, the two primary fragments spend a longer part of their evolution within the isothermal phase compared to the third secondaryy fragment. This may explain the low temperature which is associated with the secondary fragment compared to the primary fragments in the resulting triple stellar system.

\newpage

\newpage

\section{Conclusions}

In this paper, we have investigated the influence of the initial thermal state of molecular cloud cores on the formation of protostellar binary systems through gravitational fragmentation. We find that in the relatively low temperature cores in which binaries are formed(with $T_{0}$ in the range 8-10 K), the binary separation is a function of the initial thermal conditions prevailing in the molecular cloud cores. Darker and colder molecular cloud cores at $T_{0}$ = 8 K evolve into protobinary systems with small separation, whereas for slightly higher initial temperatures of 9 K or 10 K, an increase in the binary separation is observed based on the results of our models. Our quantitative analysis has therefore revealed a strong thermal sensitivity of the separation of evolving protobinary systems. Molecular cloud cores with temperature above 10 K and with small amplitude of initial azimuthal density perturbation(A = 10 $\%$)(models E and F) do not develop binary systems but form single protostars instead. Further investigation of such cores by introducing a stronger amplitude of perturbation (A = 25 $\%$), however, has shown that warm cores can indeed be forced to fragment. The evolution of the resulting binaries is apparently much slower than in the colder case, although we have not been able to follow their evolution until after pericenter passage.
We also investigated the impact of an effective cooling environment that may prolong the isothermal phase of core collapse. We find that higher values for the critical density which separates the regime of isothermal and adiabatic cloud collapse significantly affects the binary fragmentation process, in that additional secondary fragmentation has been observed in a bar connecting the primary prostellar fragments in the central part of the models. This process happens in a fairly short period of time compared to the freefall time of the cores and can give rise to the formation of multiple hierarchical protostellar systems. On the other hand, a reduction of the critical density, hence a less efficient cooling environment, suppresses fragmentation and may even lead to stable disk structures without star formation in the first place.

\section{Acknowledgements}

We thank the Institute of Space and Planetary Astrophysics (ISPA) for financially supporting this research
project. We are also grateful to the Pakistani National Center for Physics (NCP) for providing access to the computational facility that was used in this study.
We also thank the anonymous referee,  whose comments substantially improved the contents of the final version of the manuscript.

\newpage

\begin{figure*}
% \begin{minipage}{120mm}
\centering
\includegraphics[trim = 0mm 0mm 0mm 0mm, clip,width=5in]{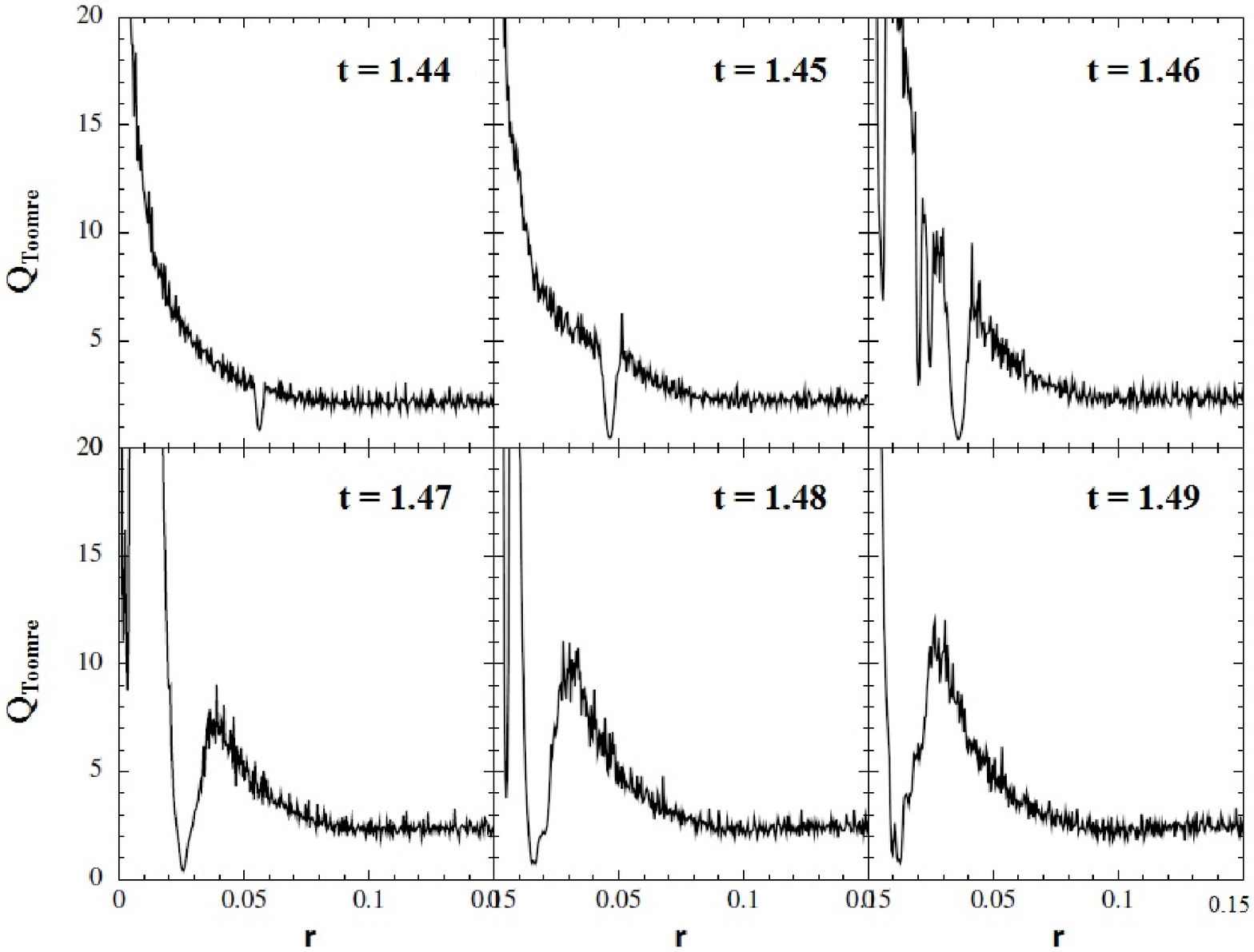}
\caption{Traces of the Toomre parameter against radial distance r measured in the equatorial plane for model F. The times for each snapshot are given in units of $t_{ff}$. }
\label{fig15}
% \end{minipage}
\end{figure*}

\begin{figure*}
% \begin{minipage}{120mm}
\centering
\includegraphics[trim = 0mm 0mm 0mm 0mm, clip, width=5in]{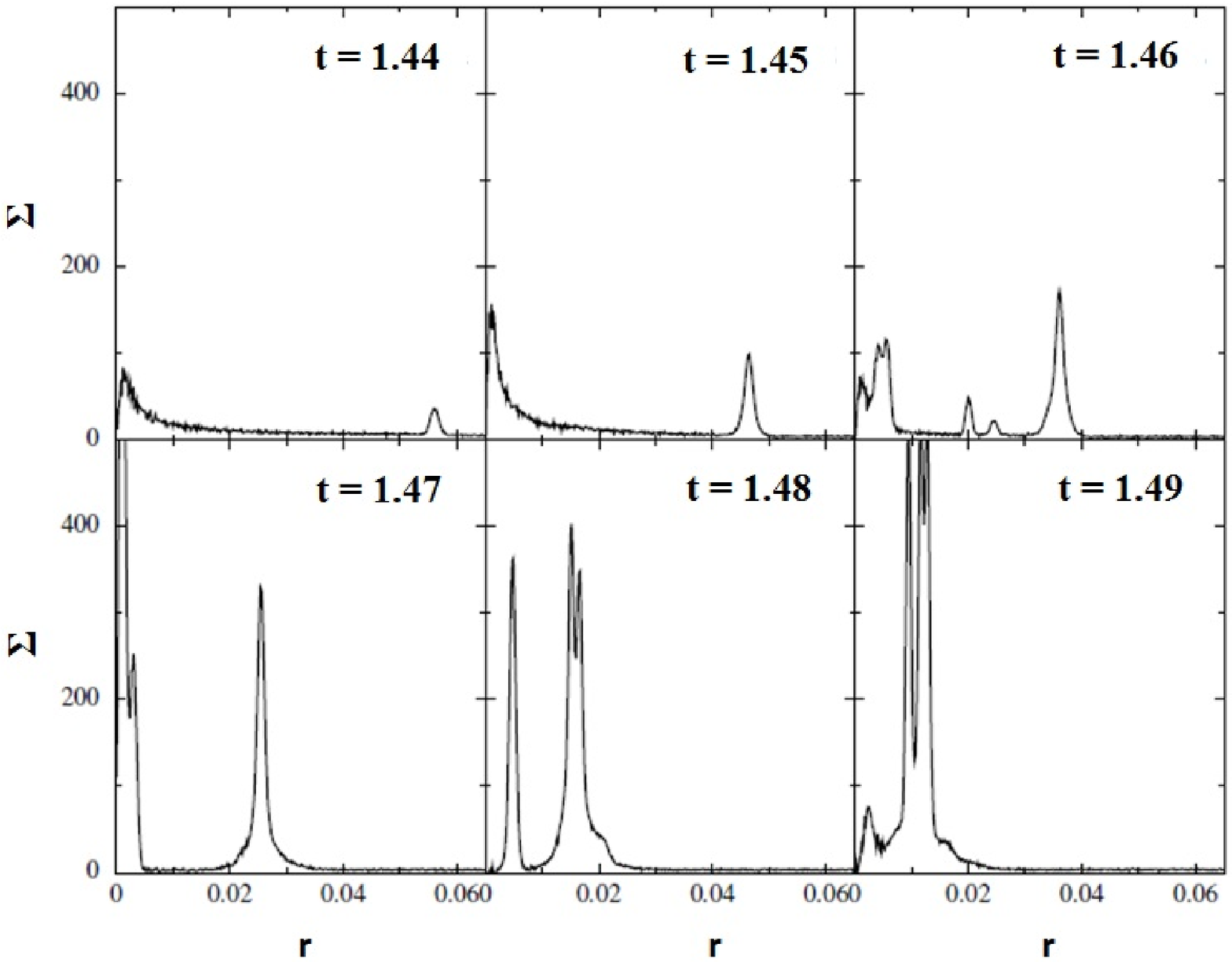}
\caption{Traces of the surface density against radial distance r measured in the equatorial plane for model F. The times for each snapshot are given in units of $t_{ff}$. }
\label{fig16}
% \end{minipage}
\end{figure*}

\begin{figure*}
% \begin{minipage}{120mm}
\centering
\includegraphics[trim = 0mm 0mm 0mm 0mm, clip, width=5in]{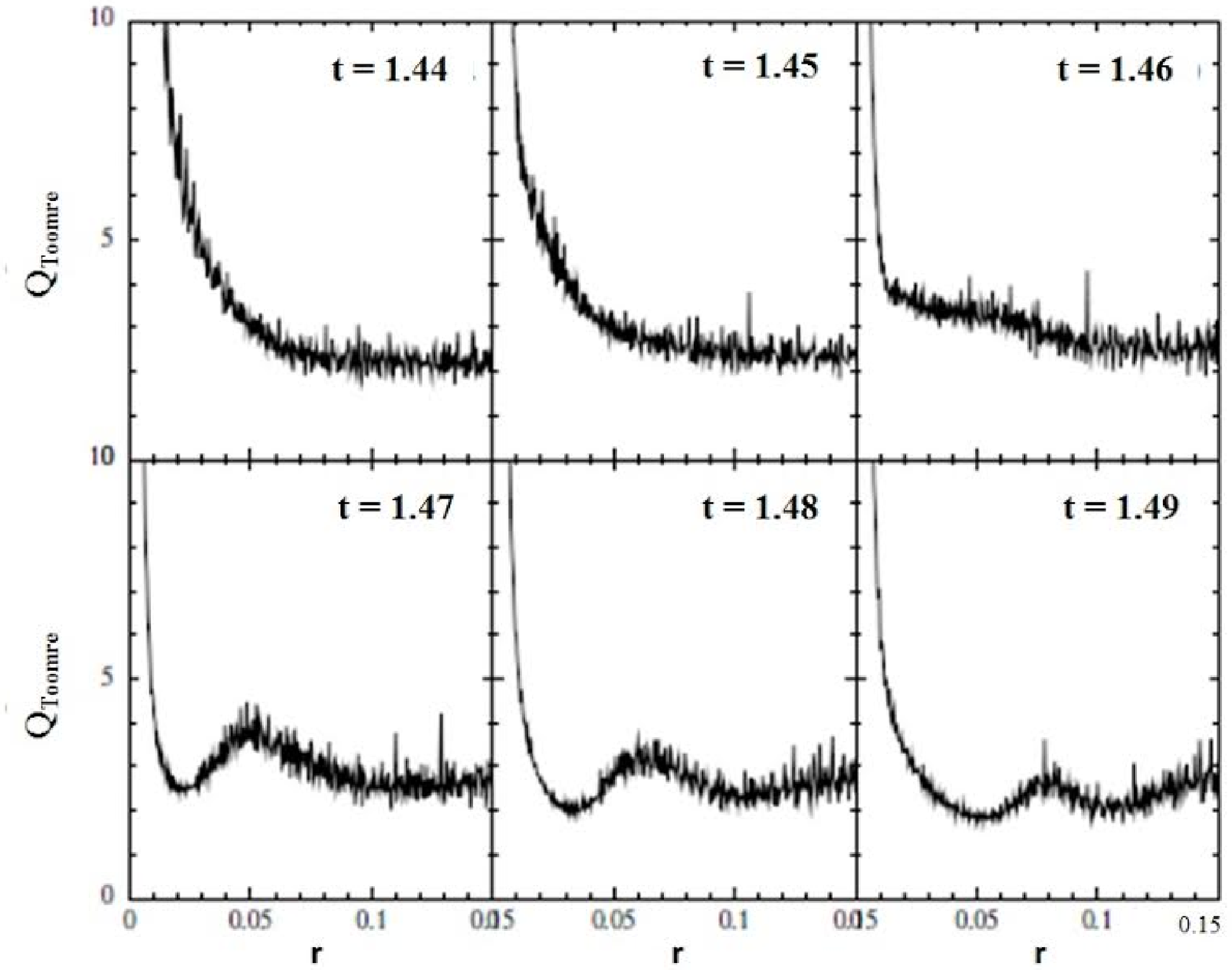}
\caption{Traces of the Toomre parameter against radial distance r measured in the equatorial plane for model G. The times for each snapshot are given in units of $t_{ff}$. }
\label{fig17}
% \end{minipage}
\end{figure*}

\begin{figure*}
% \begin{minipage}{120mm}
\centering
\includegraphics[trim = 0mm 0mm 0mm 0mm, clip, width=5in]{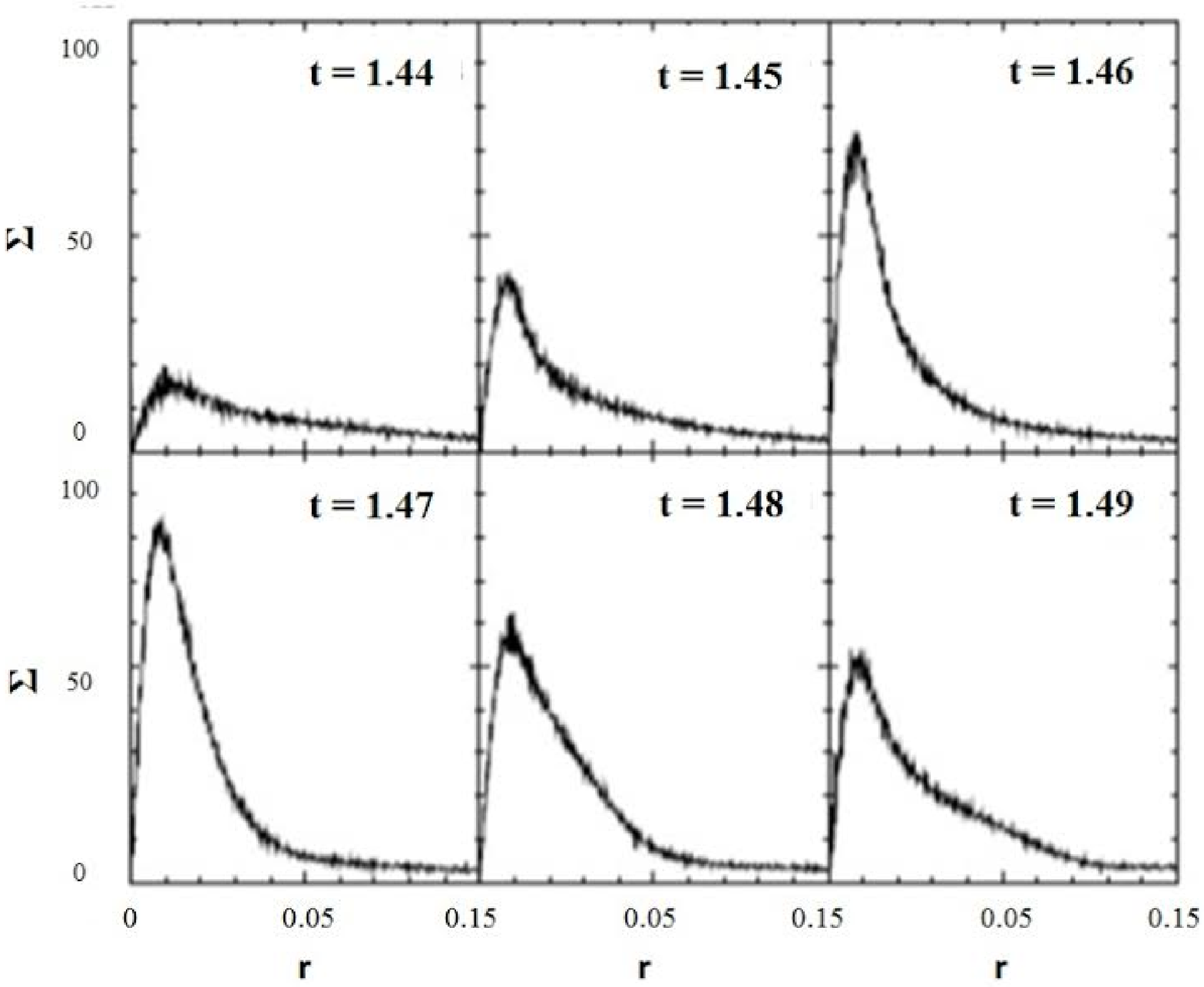}
\caption{Traces of the surface density against radial distance r measured in the equatorial plane for model G. The times for each snapshot are given in units of $t_{ff}$. }
\label{fig18}
% \end{minipage}
\end{figure*}

\begin{figure*}
% \begin{minipage}{120mm}
\centering
\includegraphics[trim = 0mm 0mm 0mm 0mm, clip, width=5in]{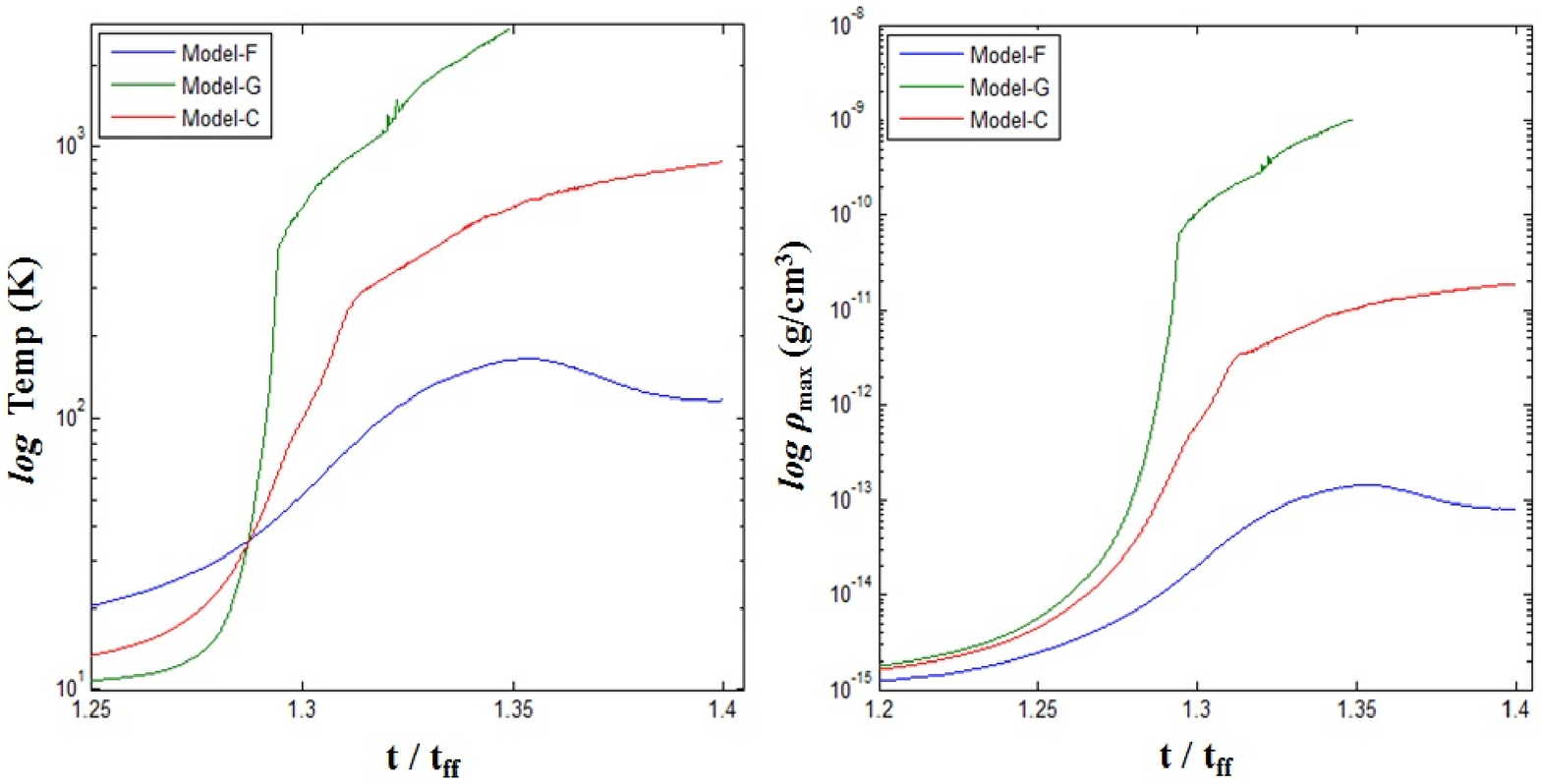}
\caption{Time evolution of the maximum temperature within  
the collapsing cloud for models E, F and G(left panel),and the time evolution 
of the maximum density for the same set of models (right panel). } 
\label{fig19}
% \end{minipage}
\end{figure*}

\begin{figure*}
% \begin{minipage}{120mm}
\centering
\includegraphics[trim = 0mm 0mm 0mm 0mm, clip, width=7in]{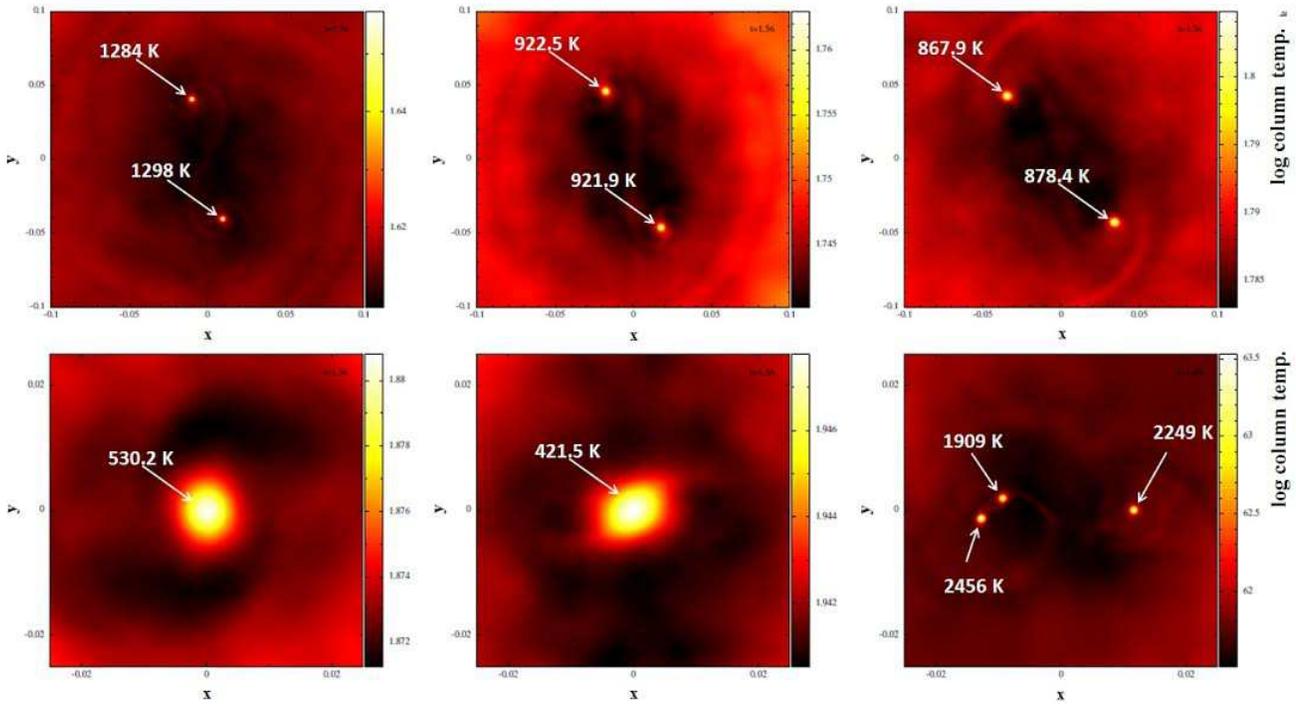}
\caption{Snapshots of the column temperature integrated along the rotation axis at the final stage of the evolution for models A, B, C (first row) and for models D, E, G (second row). 
The horizontal and vertical dimensions in the xy-plane in the first and second row are 0.1 x 0.1 and 0.025 x 0.025 in dimensionless units, respectively. 
The colour bar in each panel shows log(T) in dimensionless units. The values Arrows represent the temperature of the indicated fragment in Kelvin. }
\label{fig20}
% \end{minipage}
\end{figure*}

\label{lastpage}

\end{document}